\documentclass[10pt]{article}
\usepackage{bbm}
\usepackage[final]{graphics}
\usepackage{amsmath}
\usepackage{amsfonts,amsbsy}
\usepackage{amssymb}

\def\empile#1\over#2{\mathrel{\mathop{\kern 0pt#1}\limits_{#2}}}
\def\bs{\boldsymbol}

\newcommand{\slv}{\raise.15ex\hbox{$/$}\kern-.53em\hbox{$v$}}
\newcommand{\slF}{\raise.15ex\hbox{$/$}\kern-.53em\hbox{$F$}}
\newcommand{\slL}{\raise.15ex\hbox{$/$}\kern-.53em\hbox{$L$}}
\newcommand{\slP}{\raise.15ex\hbox{$/$}\kern-.53em\hbox{$P$}}
\newcommand{\slp}{\raise.15ex\hbox{$/$}\kern-.53em\hbox{$p$}}
\newcommand{\slq}{\raise.15ex\hbox{$/$}\kern-.53em\hbox{$q$}}
\newcommand{\slR}{\raise.15ex\hbox{$/$}\kern-.53em\hbox{$R$}}
\newcommand{\slQ}{\raise.15ex\hbox{$/$}\kern-.53em\hbox{$Q$}}
\newcommand{\slK}{\raise.15ex\hbox{$/$}\kern-.53em\hbox{$K$}}
\newcommand{\slk}{\raise.15ex\hbox{$/$}\kern-.53em\hbox{$k$}}
\newcommand{\slD}{\raise.15ex\hbox{$/$}\kern-.53em\hbox{$D$}}
\newcommand{\slC}{\raise.15ex\hbox{$/$}\kern-.53em\hbox{$C$}}
\newcommand{\slA}{\raise.15ex\hbox{$/$}\kern-.53em\hbox{$A$}}
\newcommand{\slSigma}{\raise.15ex\hbox{$/$}\kern-.53em\hbox{$\Sigma$}}
\newcommand{\slpartial}{\raise.15ex\hbox{$/$}\kern-.53em\hbox{$\partial$}}
\newcommand{\slcalP}{\raise.15ex\hbox{$/$}\kern-.63em\hbox{$\cal P$}}

\def\p{{\boldsymbol p}}
\def\q{{\boldsymbol q}}

\def\k{{\boldsymbol k}}

\def\x{{\boldsymbol x}}
\def\y{{\boldsymbol y}}

\def\u{{\boldsymbol u}}

\catcode`\@=11

\newcount\@tempcntc
\def\@citex[#1]#2{\if@filesw\immediate\write\@auxout{\string\citation{#2}}\fi
  \@tempcnta\z@\@tempcntb\m@ne\def\@citea{}\@cite{%
        \@for\@citeb:=#2\do%
    {\@ifundefined{b@\@citeb}%
        {\@citeo\@tempcntb\m@ne\@citea%
                \def\@citea{,\penalty\@m\ }{\bf ?}\@warning%
                {Citation `\@citeb' on page \thepage \space undefined}}%
        {\setbox\z@\hbox{\global\@tempcntc0\csname b@\@citeb\endcsname\relax}
     \ifnum\@tempcntc=\z@ \@citeo\@tempcntb\m@ne%
       \@citea\def\@citea{,\penalty\@m}%
       \hbox{\csname b@\@citeb\endcsname}%
     \else%
      \advance\@tempcntb\@ne%
      \ifnum\@tempcntb=\@tempcntc%
      \else\advance\@tempcntb\m@ne\@citeo%
      \@tempcnta\@tempcntc\@tempcntb\@tempcntc\fi\fi}}\@citeo}{#1}}%

\def\@citeo{\ifnum\@tempcnta>\@tempcntb\else\@citea
  \def\@citea{,\penalty\@m}%
  \ifnum\@tempcnta=\@tempcntb\the\@tempcnta\else
   {\advance\@tempcnta\@ne\ifnum\@tempcnta=\@tempcntb \else
\def\@citea{--}\fi
    \advance\@tempcnta\m@ne\the\@tempcnta\@citea\the\@tempcntb}\fi\fi}

\catcode`\@=12

\begin{document}

\title{\bf Particle production in field theories\\ 
coupled to strong external sources\\ I. Formalism and main results}
\author{Fran\c cois Gelis$^{(1)}$, Raju Venugopalan$^{(2)}$}
\maketitle
\begin{center}
\begin{enumerate}
\item Service de Physique Th\'eorique, URA 2306 du CNRS\\
  CEA/DSM/Saclay, B\^at. 774\\
  91191, Gif-sur-Yvette Cedex, France
\item Department of Physics, Bldg. 510 A,\\
Brookhaven National Laboratory,\\
  Upton, NY-11973, USA
\end{enumerate}
\end{center}

\begin{abstract}
  We develop a formalism for particle production in a field theory
  coupled to a strong time-dependent external source. An example of
  such a theory is the Color Glass Condensate.  We derive a formula,
  in terms of cut vacuum-vacuum Feynman graphs, for the probability of
  producing a given number of particles. This formula is valid to all
  orders in the coupling constant. The distribution of multiplicities
  is non--Poissonian, even in the classical approximation. We
  investigate an alternative method of calculating the mean
  multiplicity. At leading order, the average multiplicity can be
  expressed in terms of retarded solutions of classical equations of
  motion. We demonstrate that the average multiplicity at {\it
  next-to-leading order} can be formulated as an initial value problem
  by solving equations of motion for small fluctuation fields with
  retarded boundary conditions.  The variance of the distribution can
  be calculated in a similar fashion. Our formalism therefore provides
  a framework to compute from first principles particle production in
  proton-nucleus and nucleus-nucleus collisions beyond leading order
  in the coupling constant and to all orders in the source density.
  We also provide a transparent interpretation (in conventional field
  theory language) of the well known Abramovsky--Gribov--Kancheli
  (AGK) cancellations.  Explicit connections are made between the
  framework for multi-particle production developed here and the
  framework of Reggeon field theory.  \vglue 5mm
\begin{flushright}
Preprint SPhT-T06/009
\end{flushright}
\end{abstract}

\section{Introduction}
The study of multiparticle production in hadron collisions at high
energies is an outstanding problem in the study of the strong
interactions. With the advent of QCD, it was understood that semi-hard
particle production in high energy hadronic interactions is dominated
by interactions between partons having a small fraction $x$ of the
longitudinal momentum of the incoming nucleons.  In the Regge limit of
small $x$ and fixed momentum transfer squared $Q^2$ -- corresponding to
very large center of mass energies squared $s$ -- the
Balitsky-Fadin-Kuraev-Lipatov (BFKL) evolution
equation~\cite{BalitL1,KuraeLF1} predicts that parton densities grow
very rapidly with decreasing $x$.  Because this rapid growth in the
Regge limit corresponds to very large phase space densities of partons
in hadronic wave-functions, it was proposed that saturation effects may
play an important role in hadronic collisions at very high
energies~\cite{GriboLR1,MuellQ1,BlaizM1,Muell4}. Saturation effects
slow down the growth of parton densities relative to that of BFKL
evolution and may provide the mechanism for the unitarization of
cross-sections at high energies.

The large phase space density suggests that the small $x$ partons can
be described by a classical color field rather than as
particles \cite{McLerV1,McLerV2,McLerV3}. More precisely, the
McLerran-Venugopalan (MV) model proposes a dual description, whereby
small $x$ partons are described by a classical field and the large $x$
partons act as color sources for the classical field.  The original
model considered a large nucleus containing a large number of large
$x$ partons (at least $3A$ valence quarks where $A$ is the atomic
number of the nucleus). In this limit, they produce a strong color
source and one has to solve the full classical Yang-Mills equations to
find the classical field. This procedure properly incorporates the
recombination interactions that are responsible for gluon saturation.
In the MV model, the large $x$ color sources are described by a
Gaussian statistical distribution~\cite{McLerV1,Kovch1}. A more
general form of this statistical distribution, for $SU(N_c)$ gauge
theories, valid for large $A$ and moderate $x$, is given in
Ref.~\cite{JeonV1,JeonV2}.

The separation between large $x$ and small $x$, inherent to the dual
description of the MV model, is somewhat arbitrary and has been
exploited to derive a renormalization group (RG) equation, the JIMWLK
equation
\cite{JalilKMW1,JalilKLW1,JalilKLW2,JalilKLW3,JalilKLW4,IancuLM1,IancuLM2,FerreILM1}.
This functional RG equation describes the change in the 
statistical distribution of color sources with $x$. It can be
expressed as an infinite hierarchy of evolution equations for
correlators \cite{Balit1}, and has a useful (and tremendously simpler)
large $N_c$ and large $A$ mean-field approximation \cite{Kovch3},
known as the Balitsky-Kovchegov equation. This general framework is
often referred to as the Color Glass Condensate
(CGC)~\cite{McLer1,IancuLM3,IancuV1}.

The discussion above refers to the properties of hadronic wave
functions at high energies.  To compute particle production in the CGC
framework, (a) one must find the distribution of color sources in the
projectiles, either by solving the JIMWLK equation or by using a model
such as the MV model, and (b) one must calculate the production of the
relevant particles from the sources. In this paper, we will assume
that the source distributions are known and shall focus on particle
production in the presence of these sources.

At leading order, one knows how to compute the production of quarks
and gluons from strong color sources.  Inclusive gluon production is
obtained by solving the classical Yang-Mills equations for two color
sources moving at the speed of light in opposite
directions~\cite{KovneMW1,KovneMW2,KovchR1}. This problem has been
solved numerically in
\cite{KrasnV4,KrasnV1,KrasnV2,KrasnNV1,KrasnNV2,Lappi1} for the
boost-invariant case. A first computation for the boost non-invariant
case has also been performed recently~\cite{RomatV1}. The multiplicity
of quark-pairs is computed from the quark propagator in the background
field of \cite{KrasnV4,KrasnV1,KrasnV2,KrasnNV1,KrasnNV2,Lappi1} -- it
has been studied numerically in \cite{GelisKL1,GelisKL2}.

These results summarize the state of the art in studies of particle
production from strong classical color sources. In the present paper,
we will systematically examine particle production in a field theory
coupled to an external time-dependent classical source. In particular,
we will consider how one computes results beyond the above mentioned
leading order results.  To avoid the technical complications of gauge
theories, we shall focus our discussion on the theory of a real scalar
field. The complexities of QCD, such as gauge invariance, are of
course extremely important. However, several of the lessons gained
from the simpler theory considered here should also apply to studies
of particle production in QCD. These will be applied in follow ups to
this paper.

This paper is organized as follows. We begin, in the next section, by
setting up  the model.  We recall well known results for
counting powers of the coupling constant and of the powers of $\hbar$;
in particular we discuss how these are modified for field theories
with strong external sources.  We discuss the vacuum-vacuum diagrams
that will play a crucial role in the rest of the paper. We show
clearly that the sum of the vacuum-vacuum diagrams is not a trivial
phase (as is the case for field theories in the vacuum) but gives a
non-trivial contribution to the probability in theories where the
field is coupled to a time-dependent source.

In section \ref{sec:distribution}, we use Cutkosky's cutting rules
to derive a formula for the probability $P_n$ of producing $n$
particles, in terms of cut vacuum-vacuum diagrams. This formula, for
theories with strong sources, is new and is valid to all orders in the
coupling constant.  While the formula at this stage is not useful for
explicit computations, it shows unambiguously that the distribution of
the multiplicities of produced particles is not Poissonian, even in
the classical approximation. We also derive a compact formula for the
generating function $F(x)$ of the probabilities $P_n$.

In section \ref{sec:keldysh}, we discuss in detail the calculation of
the first moment of the distribution, namely, the average multiplicity
$\big<n\big>=\sum_n n\,P_n$. We obtain a formula which is valid to all
orders for $\big<n\big>$. We work it out explicitly at leading and
next-to-leading order. It is well known that the leading order
expression can be expressed in terms of classical fields that are
computed by solving partial differential equations, with retarded
boundary conditions, for the classical fields. What is remarkable and
a major result of this work is the fact that one can similarly compute
the multiplicity at NLO from the small fluctuation fields (in the
presence of the classical background field) with simple initial
conditions at $x_0=-\infty$, i.e.  in terms of partial differential
equations with retarded boundary conditions. {\it This result makes
next-to-leading order computations of the average multiplicity
feasible even in theories with strong external sources. It has
possibly very significant ramifications for heavy ion collisions.}

The results in section~\ref{sec:distribution} and \ref{sec:keldysh}
are related in section \ref{sec:agk}. The Abramovsky-Gribov-Kancheli
cancellations \cite{AbramGK1}, originally formulated in the context of
reg\-geon field theory models, play an important role in relating the
inclusive particle multiplicities in \ref{sec:keldysh} to moments of
the probability distributions in \ref{sec:distribution}.  
In section \ref{sec:agk}, we provide a ``dictionary'' to map the
original AGK discussion formulated in the language of cut reggeons
into our language of cut vacuum-vacuum diagrams in field theories with
external sources. The fact that the AGK cancellations require a
Poissonian distribution of cut reggeons is seen to follow naturally if
one identifies reggeons (cut and uncut) with connected vacuum-vacuum
sub-diagrams. This identification further leads us to the conclusion
that there are further cancellations that are not visible when the
discussion takes place at the level of reggeons. These additional
cancellations simplify the calculation of the multiplicity
$\big<n\big>$ as well as that of higher moments. We end this
section by a general formula that gives the variance of the
distribution of produced particles.

In the final section, we provide a summary of the new results derived
in this paper and of their possible ramifications for QCD at high
energies. The paper contains four appendices. In appendix
\ref{app:SK}, we give a very brief reminder of the Schwinger-Keldysh
formalism. In appendix~\ref{app:LS}, we derive the solution of the
Lippmann-Schwinger equation (\ref{eq:LS}).  In
appendix~\ref{app:inv-agk}, we discuss a proof demonstrating that the
only distributions of cut reggeons that lead to AGK cancellations are
Poisson distributions. In appendix~\ref{app:cancellations}, we discuss
details of the cancellations giving rise to the identity in
eq.~(\ref{eq:agk-rhs2}).

\section{General framework}
\label{sec:framework}
In this section, we will set the groundwork for the discussion in
subsequent sections. Much of the material here is textbook fare and
can be found, for instance, in refs.~\cite{ItzykZ1,PeskiS1}. However,
there are subtleties that distinguish the discussion of field theories
in the presence of external sources to discussions of field theories
in vacuum. These are emphasized throughout, especially in
section~\ref{sec:vacuum}.

\subsection{Model}
\label{sec:model}
The model, for all the discussions in this paper, will be the theory
of a real scalar field $\phi$, with $\phi^3$ self-interactions and
with the field coupled to an external source $j(x)$. This simple
scalar theory captures many of the key features of multi-particle
production in QCD, while avoiding the complications arising from
keeping track of the internal color degrees of freedom and the
necessity of choosing a gauge in the latter.

Without further ado, the Lagrangian density in our model is
\begin{equation}
{\cal L}\equiv\frac{1}{2}\partial_\mu\phi\,\partial^\mu\phi 
-\frac{1}{2}m^2\phi^2-\frac{g}{3!}\phi^3 +j\phi\; .
\label{eq:lagrangian}
\end{equation}
The reader should note that the coupling $g$ in this theory is
dimensionful, with dimensions of the mass; the theory therefore, in 4
dimensions, is super renormalizable.  When we draw the Feynman
diagrams for this theory, we will denote the source by a black dot.
Keeping in mind applications of our results to the description of
hadronic collisions in the Color Glass Condensate framework, one may
think of $j(x)$ as the scalar analog of the sum of two source terms
$j(x)=j_1(x)+j_2(x)$ corresponding respectively to the color currents
of the two hadronic projectiles.

\subsection{Power counting}
\label{sec:counting}
Power counting is essential in understanding the perturbative
expansion of amplitudes in this model.  For theories with weak
sources, where $j$ is parametrically of order $1$, the leading
contributions would come from the diagrams with the least number of
vertices. We are here interested in theories (such as the CGC) where
the sources are strong.  By strong, we mean specifically that the
$j\,\phi$ term is of the same order as the other terms in the
Lagrangian density. This implies that $g\,j(x)$ is of order unity, as
opposed to $j(x)\sim 1$ itself. Therefore, as discussed further below,
independent powers of $g$ in any given diagram arise only if the
vertex is not connected to a source. In this power counting scheme,
the leading diagrams can have negative powers of $g$. We will
subsequently also discuss the power counting in $\hbar$.

\subsubsection{Powers of the coupling}
For strong sources, $g\,j\sim 1$, attaching new sources to a given
diagram does not change its order. There are therefore an infinity of
diagrams contributing at a given order in $g$. For instance, even the
classical approximation of this model, which involves only tree
diagrams, resums an infinite number of contributions. Field theories
coupled to a strong source are therefore non-trivial and exhibit rich
properties {\sl even if the coupling $g$ is very small}.

To make this discussion more explicit, consider a generic
simply connected diagram containing $n_{_E}$ external lines, $n_{_I}$
internal lines, $n_{_L}$ loops, $n_{_J}$ sources and $n_{_V}$
vertices. These parameters are not independent, but are related by the
equations,
\begin{eqnarray}
&&
3n_{_V}+n_{_J}=n_{_E}+2n_{_I}\; ,
\nonumber\\
&&
n_{_L}=n_{_I}-n_{_V}-n_{_J}+1\; .
\end{eqnarray}
Thanks to these two relations, the order of this diagram
is
\begin{equation}
g^{n_{_V}} j^{n_{_J}}
=
g^{n_{_E}+2(n_{_L}-1)}\big(gj\big)^{n_{_J}}\; .
\end{equation}
As explained previously, when the source is as strong as $1/g$, the
last factor is irrelevant; the order of the diagram depends only on
its number of external legs and number of loops. Note also that the
above formulas are only valid for a simply connected diagram: they
must be applied independently to each disconnected subdiagram of a
more complicated diagram. Note in particular that a simply connected
tree diagram ($n_{_L}=0$) with no external legs ($n_{_E}=0$) attached
to strong sources is of order $g^{-2}$.

\subsubsection{Powers of $\hbar$}
It is also interesting to discuss the order in $\hbar$ of a given
diagram. Recall that $\hbar$ has the dimension of
an action. Therefore, 
\begin{equation}
\frac{S}{\hbar}
=
\int d^4x\;
\left[
-\frac{1}{2
}\phi\frac{G_0^{-1}}{\hbar}\phi
-\frac{g}{3!\hbar}\phi^3
+\frac{j}{\hbar}\phi
\right]\; ,
\end{equation}
where $G_0$ is the free propagator. {}From this equation, it is clear
that the following mnemonic can be applied to keeping track of powers
of $\hbar$ in the perturbative expansion:
\begin{itemize}
\item[(i)] a propagator brings a power of $\hbar$,
\item[(ii)] a vertex brings a power of $1/\hbar$,
\item[(iii)] a source $j$ brings a power of $1/\hbar$.
\end{itemize}
The order in $\hbar$ of the generic diagram in the previous
sub-section is therefore
\begin{equation}
\hbar^{n_{_E}+n_{_I}-n_{_V}-n_{_J}}=\hbar^{n_{_E}+n_{_L}-1}\; .
\end{equation}
We thus recover the well-known result that the order in $\hbar$, for a
fixed number of external legs, depends only on the number of loops.
For the purposes of this paper, it is important to note that the
result does not depend on the number of insertions of the classical
source in the diagram. Clearly this implies that arbitrarily
complicated diagrams can be of the same order in $\hbar$. In the rest
of this paper, we do not keep track of the powers of $\hbar$, and
simply set $\hbar=1$.

\subsection{Vacuum--vacuum diagrams}
\label{sec:vacuum}
The forthcoming discussion in the next section, of the multiplicity
distributions of produced particles, will be formulated in terms of
``vacuum-vacuum'' diagrams in the presence of strong external sources.
Vacuum--vacuum diagrams are simply Feynman diagrams without any
external legs~\cite{ItzykZ1}. They contribute to the vacuum-to-vacuum
transition amplitude $\big<0_{\rm out}\big|0_{\rm in}\big>$ -- hence
the name. The perturbative expansion of any amplitude
with external legs generates these vacuum--vacuum diagrams as
disconnected factors.

When one considers a field theory in the vacuum (no external source:
$j=0$), one usually disregards vacuum--vacuum graphs because their sum
is a pure phase that does not contribute to transition probabilities.
That the sum of vacuum-vacuum diagrams, in this case, is a pure phase
follows from the fact that no particles can be created if the initial
state is the vacuum. The probability that the vacuum stays the vacuum
is unity and the corresponding amplitude must be a pure phase.

The situation is very different in a theory, like the one under
consideration, where the fields are coupled to an external
source\footnote{For example, in the Color Glass Condensate (CGC)
framework, incoming ha\-dro\-nic states are described by a pair of
classical color sources to which the gauge fields couple. At high
energies, such a separation is dictated by the kinematics. The
resulting effective theory, the CGC, greatly simplifies the
description of the initial
state~\cite{Muell4,McLer1,IancuLM3,IancuV1}.}. Assume that this
external source is such (time-dependent for instance) that an
amplitude from the vacuum to a populated state is non-zero, namely,
\begin{equation}
\exists\beta\not=0\;,\quad
\big<\beta_{\rm out}\big|0_{\rm in}\big>\not=0\; .
\end{equation}
{}From unitarity,
\begin{equation}
\sum_\beta \Big|\big<\beta_{\rm out}\big|0_{\rm in}\big>\Big|^2=1\; ,
\end{equation}
we conclude that
\begin{equation}
\Big|\big<0_{\rm out}\big|0_{\rm in}\big>\Big|^2 <1\; .
\end{equation}
Therefore, the amplitude $\big<0_{\rm out}\big|0_{\rm in}\big>$ cannot
be a pure phase.

It can be expressed as~\cite{ItzykZ1}
\begin{equation}
\big<0_{\rm out}\big|0_{\rm in}\big>\equiv e^{i{\cal V}[j]} \, ,
\end{equation}
where $i{\cal V}[j]$ denotes\footnote{The factor $i$ is purely
  conventional here. It has been introduced for later convenience when
  we discuss cutting rules.} compactly the sum of the connected
  vacuum-vacuum diagrams in the presence of an external source $j$,
  \setbox1=\hbox to
  7cm{\resizebox*{7cm}{!}{\includegraphics{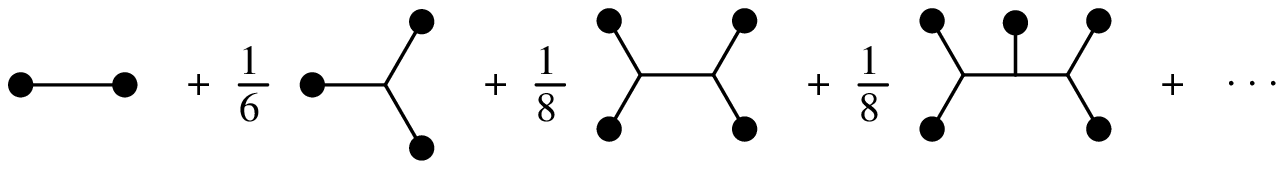}}}
\begin{equation}
i{\cal V}[j]\equiv i\sum_{\rm conn}V = \quad\;\raise -3.5mm\box1
\label{eq:Vconn}
\end{equation}

In the diagrammatic expansion on the right hand side, we have
  represented only tree diagrams, but of course $i{\cal V}[j]$
  contains diagrams at any loop order.  The number preceding each
  diagram is its symmetry factor.  Note finally that because external
  lines are absent, the power counting of the previous subsection
  tells us that vacuum-vacuum diagrams come with powers that are given
  by
\begin{equation}
g^{2(n_{_L}-1)}\big(gj\big)^{n_{_J}}\; .
\end{equation}
Therefore, in the presence of strong sources, $j\sim g^{-1}$, connected
tree vacuum-vacuum diagrams are all of order $g^{-2}$, the 1-loop
vacuum-vacuum diagrams are all of order $1$, and so on. 

It is the squared modulus of the vacuum--vacuum amplitude that
contributes to the transition probability. This is simply $\exp
\big(-2\,{\rm Im}\,{\cal V}[j]\big)$. In the next section, we
will discuss a method for computing $2\,{\rm Im}\,{\cal V}[j]$
directly.

\section{Multiplicity distribution from\\
vacuum-vacuum diagrams}
\label{sec:distribution}
We will begin this section with a discussion of the cutting rules for
vacuum-vacuum graphs in theories with external sources. These rules,
for instance, can be employed to compute the imaginary part of the
${\cal V}[j]$ introduced in eq.~(\ref{eq:Vconn}).  We will then
present an intuitive discussion of how the probability for producing
$n$-particles can be expressed in terms of cut vacuum-vacuum
diagrams. We shall demonstrate the importance of the vacuum-vacuum
factors in preserving unitarity for a theory with strong external
sources. A generating function is introduced to compute moments of the
distribution of probabilities; we will show that the distribution of
probabilities is non--Poissonian, even for classical tree level
graphs.  In the final sub-section, we will present a formal expression
of the intuitive arguments developed in the previous sub-sections.

\subsection{Cutting rules}
\label{sec:cutting}
Let us first consider the analog of the Cutkosky rules \cite{Cutko1}
in our model.  These will also prove useful for our discussion in the
following section where we develop an alternative method to compute
multiplicity distributions. One starts by decomposing the Feynman
(time-ordered) free propagator in two pieces according to the
time-ordering of the two endpoints\footnote{The superscript 0
indicates {\bf free} propagators.}
\begin{equation}
G^0_{_F}(x,y)\equiv \theta(x^0-y^0)G^0_{-+}(x,y)+\theta(y^0-x^0)G^0_{+-}(x,y)\; .
\end{equation}
This equation defines the propagators $G^0_{-+}$ and $G^0_{+-}$. The
notations for these objects have been chosen in analogy with the
Schwinger-Keldysh formalism (see \cite{Schwi1,Keldy1}, and the brief
summary in appendix \ref{app:SK}), for reasons that will become clear
in section \ref{sec:keldysh}. Pursuing this analogy, the Feynman
propagator $G^0_{_F}$ is also denoted as $G^0_{++}$.  We introduce as
well the (anti--time-ordered) propagator $G^0_{--}$ defined as
\begin{equation}
G^0_{--}(x,y)\equiv \theta(x^0-y^0)G^0_{+-}(x,y)+\theta(y^0-x^0)G^0_{-+}(x,y)\; .
\end{equation}
Note that the four propagators so defined are not independent; they
are related by the identity,
\begin{equation}
G^0_{++}+G^0_{--}=G^0_{-+}+G^0_{+-}\; .
\end{equation}
In addition to these new propagators, we will consider two kinds of
vertices, of type $+$ or $-$. A vertex of type $+$ is the ordinary
vertex and appears with a factor $-ig$ in Feynman diagrams. A vertex
of type $-$ is the opposite of a $+$ vertex, and its Feynman rule is
$+ig$. Likewise, for insertions of the source $j$, insertions of type
$+$ appear with the factor $+ij(x)$ while insertions of type $-$
appear instead with $-ij(x)$.  The motivation for introducing these
additional vertices and sources will become clearer shortly.

For each Feynman diagram $iV$ containing only ``+'' vertices and
sources (denoted henceforth as $iV_{\{+\cdots+\}}$) contributing to
the sum of connected vacuum-vacuum diagrams, define a corresponding
set of diagrams $iV_{\{\epsilon_i\}}$ by assigning the symbol
$\epsilon_i$ to the vertex $i$ of the original diagram (and connecting
a vertex of type $\epsilon$ to a vertex of type $\epsilon^\prime$ with
the propagator $G^0_{\epsilon\epsilon^\prime}$). Each $\epsilon_i$ can
be either of the ``+'' or ``-'' type. The generalized set of diagrams
includes $2^n$ such diagrams if the original diagram had $n$ vertex
and sources. These diagrams obey the so-called ``largest time
equation''~\cite{t'HooV1,Veltm1}. If one examines these diagrams
before the times at the vertices and sources have been integrated out,
and assumes that the vertex or source with the largest time is
numbered $i$, it is easy to prove that
\begin{equation}
iV_{\{\cdots\epsilon_i\cdots\}}+iV_{\{\cdots-\epsilon_i\cdots\}}=0\; ,
\end{equation}
where the dots denote the exact same configuration of $\epsilon$'s in
both terms, for the vertices and sources that do not carry the largest
time. This equation generalizes immediately to the constraint, 
\begin{equation}
\sum_{\{\epsilon_i\}}iV_{\{\epsilon_i\}}=0\; ,
\label{eq:pm-constraint}
\end{equation}
where the sum is extended to the $2^n$ possible configurations of the
$\epsilon$'s in the Feynman diagram of interest~\footnote{For a recent
  application of largest time relations to QCD, we refer the reader to
  Ref.~\cite{VamanY1}.}.

This well known identity, being true for arbitrary times and positions
of the vertices and sources, is also valid for the Fourier transforms
of the corresponding diagrams. To see this, we first write down the
explicit expressions for the Fourier transforms of the propagators
$G^0_{\pm\pm}$
\begin{eqnarray}
&&
G^0_{++}(p)=\frac{i}{p^2-m^2+i\epsilon}\;,\quad
G^0_{--}(p)=\frac{-i}{p^2-m^2-i\epsilon}\; ,
\nonumber\\
&&
G^0_{-+}(p)=2\pi\theta(p^0)\delta(p^2-m^2)\; ,\quad
G^0_{+-}(p)=2\pi\theta(-p^0)\delta(p^2-m^2)\; .
\label{eq:Fourier-props}
\end{eqnarray}
$G^0_{--}(p)$ is the complex conjugate of $G^0_{++}(p)$.  Using these
relations, one can show that
\begin{equation}
iV_{\{+\cdots+\}}+iV_{\{-\cdots-\}}
=iV_{\{+\cdots+\}}+\left(iV_{\{+\cdots+\}}\right)^*
=-2\,{\rm Im}\,V\; .
\end{equation}
We can therefore rewrite the imaginary part of the original Feynman
diagram $V$ as
\begin{equation}
2\,{\rm Im}\,V = \sum_{\{\epsilon_i\}^\prime}
iV_{\{\epsilon_i\}}\; ,
\label{eq:cut1}
\end{equation}
where the prime in the sum indicates that the two terms where all the
vertices and sources are of type $+$ or all of type $-$ are excluded
from the sum.

For a given term in the right hand side of the above formula, one can
divide the diagram in several disconnected subgraphs, each containing
only $+$ or only $-$ vertices and sources\footnote{Note that such
disconnected subgraphs can exist only if they contain at least one
external source.  They would otherwise be forbidden by energy
conservation.}. The $+$ regions and $-$ regions of the diagram are
separated by a ``cut'' -- each diagram in the r.h.s. of
eq.~(\ref{eq:cut1}) is therefore a ``cut vacuum-vacuum diagram".  At
tree level, the first terms generated by these cutting rules (applied
to compute the imaginary part of the sum of connected vacuum-vacuum
diagrams) are \setbox1=\hbox to
8cm{\resizebox*{8cm}{!}{\includegraphics{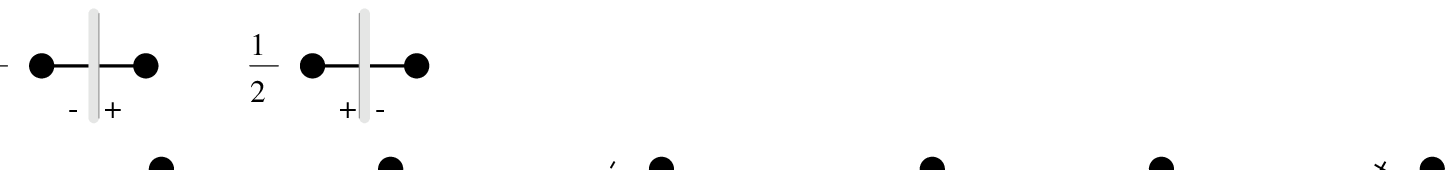}}}
\begin{eqnarray}
&&
2\,{\rm Im}\,{\cal V}[j]=\quad\;\raise -24mm\box1\nonumber\\
&&\qquad\qquad\qquad\;\, + \cdots
\label{eq:a}
\end{eqnarray}
The $+$ and $-$ signs adjacent to the grey line in each diagram here
indicate the side on which the set of $+$ and $-$ vertices is located.
As one can see, there are cuts intercepting more than one propagator.
In the following, a cut going through $r$ propagators will be called a
``$r$-particle cut''.

It is important to note that cut vacuum-vacuum diagrams would be zero
in the vacuum because energy cannot flow from from one side of the cut
to the other in the absence of external legs.  This constraint is
removed if the fields are coupled to {\bf time-dependent} external
sources. Thus, in this case, cut vacuum-vacuum diagrams, and therefore
the imaginary part of vacuum-vacuum diagrams, differ from zero.

\subsection{Probability of producing $n$ particles}
\label{sec:Pn}
All transition amplitudes contain the factor $\exp\big(i{\cal
V}[j]\big)$, which, in the squared amplitudes, transforms to the
fac\-tor $\exp \big(-2\,{\rm Im}\,{\cal V}[j]\big)$. The power
counting rules derived earlier tell us that the sum of all the
connected vacuum-vacuum diagrams starts at order\footnote{From now on,
we shall not write explicitly the dependence in the combination $gj$
because we shall assume the fields are strong and this combination is
of order unity.  All the coefficients in the expressions we write in
this section depend implicitly on $gj$.} $g^{-2}$. Hence, we will
write
\begin{equation}
{i({\cal V}[j]-{\cal V}^*[j])} = 2\,{\rm Im}\,{\cal V}[j]
 \equiv \frac{a}{g^2}\; ,
\label{eq:im-V}
\end{equation}
where $a$ denotes a series in $g^{2n}$ that starts at order
$n=0$;  the coefficients of this series are functions of $gj$.
Thus, the vacuum-to-vacuum transition probability is $\exp(-a/g^2)$.
The simplest tree diagrams entering in $a/g^2$ were displayed in
eq.~(\ref{eq:a}).

We now turn to the probabilities for producing $n$ particles. For now,
our discussion is more intuitive than rigorous, with precise
definitions to be introduced later in the section (in \ref{sec:Pn-1}).
Besides the overall factor $\exp\big(i{\cal V}[j]\big)$, the
transition amplitude from the vacuum state to a state containing one
particle is the sum of all the Feynman diagrams with one external
line. These diagrams start at order $g^{-1}$. The probability to
produce one particle from the vacuum can therefore be parameterized as
\begin{equation}
P_1 = e^{-a/g^2} \; \frac{b_1}{g^2}\; ,
\end{equation}
where $b_1$ is, like $a$, a series in $g^{2n}$
that starts at $n=0$. $b_1/g^2$ can be obtained by performing a 1-particle cut 
through vacuum-vacuum diagrams. In other words, $b_1/g^2$ is one
of the contributions that one can find in $a/g^2$. Diagrammatically,
$b_1/g^2$ starts at tree level with
\setbox1=\hbox to 8cm{\resizebox*{8cm}{!}{\includegraphics{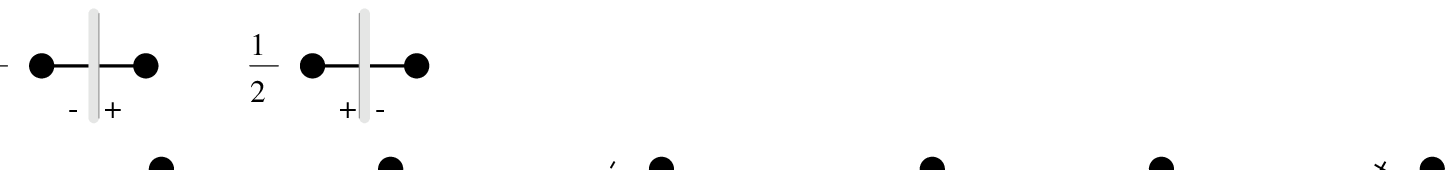}}}
\begin{eqnarray}
\frac{b_1}{g^2}&=&\quad\;\raise -13mm\box1 \nonumber \\
&+&\cdots
\label{eq:b1}
\end{eqnarray}

Consider now the probability $P_2$ for producing two particles from
the vacuum. There is an obvious contribution to this probability that
is obtained simply by squaring the $b_1/g^2$ piece of the probability
for producing one particle. This term corresponds to the case where
the two particles are produced independently from one another. But two
particles can also be produced correlated to each other. This
``correlated" contribution to $P_2$ must come from a 2-particle cut
through connected vacuum-vacuum diagrams. Let us represent this
quantity as $b_2/g^2$. Diagrammatically, $b_2/g^2$ is a series whose
first terms are \setbox1=\hbox to
8cm{\resizebox*{8cm}{!}{\includegraphics{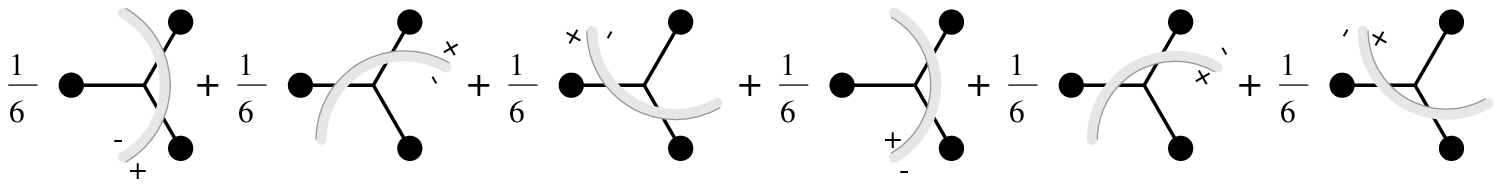}}}
\begin{eqnarray}
\frac{b_2}{g^2}&=&\quad\;\raise -4.5mm\box1 \nonumber \\
&+&\cdots
\label{eq:b2}
\end{eqnarray}
The net probability, from correlated and uncorrelated production, of
two particles can therefore be represented as
\begin{equation}
P_2= e^{-a/g^2} \; \left[\frac{1}{2!}\frac{b_1^2}{g^4} 
+ \frac{b_2}{g^2}\right]\; ..
\end{equation}
The prefactor $1/2!$ in front of the first term is a symmetry factor which is 
required because the two particles in the final state are undistinguishable.

Let us further discuss the case of three particle production before
proceeding to the general case. One (``uncorrelated") term will be the
cube of $b_1/g^2$ (preceded by a symmetry factor $1/3!$). A
combination $b_1 b_2 / g^4$ will also appear, corresponding to the
case where two of the particles are produced in the same subdiagram,
and the third is produced independently. Finally, there is an
``intrinsic" three particle production probability corresponding to
the three particles produced in the same diagram.  We shall represent
this contribution by $b_3/g^2$. More precisely, $b_3/g^2$ is the sum
of all 3-particle cuts in $a/g^2$. Diagrammatically, some of the
simplest terms in $b_3/g^2$ are \setbox1=\hbox to
5.3cm{\resizebox*{5.3cm}{!}{\includegraphics{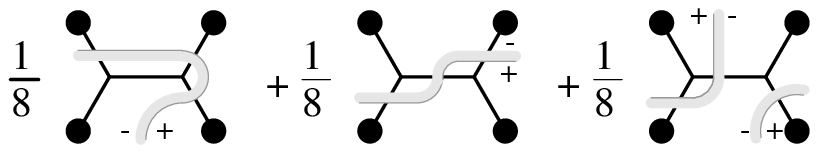}}}
\begin{eqnarray}
\frac{b_3}{g^2}&=&\quad\;\raise -3.5mm\box1 \nonumber \\
&+&\cdots
\label{eq:b3}
\end{eqnarray}
(Only a few terms have been represented at this order, due to the
large number of possible permutations of cuts across the various
legs.) The probability of producing three particles from the vacuum is
then given by
\begin{equation}
P_3 = e^{-a/g^2} \; \left[\frac{1}{3!}\frac{b_1^3}{g^6} 
+\frac{b_1b_2}{g^4}+ \frac{b_3}{g^2}\right]\; .
\end{equation}

The previous examples can be generalized to obtain an expression for
the production of $n$ particles, for any $n$. It reads
\begin{equation}
P_n=e^{-a/g^2} \; \sum_{p=0}^n \frac{1}{p!}
\sum _{\alpha_1+\cdots+\alpha_p=n}
\frac{b_{\alpha_1}\cdots b_{\alpha_p}}{g^{2p}}\; .
\label{eq:Pn}
\end{equation}
In this formula, $p$ is the number of disconnected subdiagrams
producing the $n$ particles, and $b_r/g^2$ denotes the contribution to
the probability of the sum of all $r$-particle cuts through the
connected vacuum-vacuum diagrams.  This formula gives the probability
of producing $n$ particles to all orders in the coupling $g$ in a
field theory with strong external sources~\footnote{Recall also that,
in addition to the factors of $g^2$ that appear explicitly in the
formula, the terms $a, b_r$ are series in $g^{2n}$ where, further, the
coefficients of the series are functions of $gj\sim 1$}.  To the best
of our knowledge, this formula is new. Even though the dynamical
details are not specified, its compact analytical form can be used to
derive very general results for multi-particle production. These
results will be discussed extensively in the section on AGK rules.
Indeed, their generality follows from the fact that the dynamical
details are not specified. A more rigorous derivation of this formula
is presented in section \ref{sec:Pn-1}.

\subsection{Unitarity}
In our framework, the statement of unitarity is simply that the sum of the
probabilities $P_n$ be unity, namely, 
\begin{equation}
\sum_{n=0}^{\infty}P_n=1\; .
\end{equation}
 From eq.~(\ref{eq:Pn}), it is a simple matter of algebra to show that the 
 l.h.s. above is given by 
\begin{equation}
\sum_{n=0}^{\infty}P_n = e^{-a/g^2} \;
\exp\left[\frac{1}{g^2}\sum_{r=1}^{\infty}b_r\right]\; .
\end{equation}
Unitarity therefore requires that 
\begin{equation}
a=\sum_{r=1}^{\infty}b_r\; .
\label{eq:unitarity-cond}
\end{equation}
This relationship between $a$ and the $b_r$'s is in fact an identity
following directly from their respective definitions. Indeed, recall
that (see eq.~(\ref{eq:im-V})) $a/g^2$ is equal to twice the imaginary
part of the sum of connected vacuum-vacuum diagrams. Therefore,
$a/g^2$ is the sum of all the possible cuts through these connected
vacuum-vacuum diagrams. On the other hand, $b_r/g^2$ was defined as
the subset of cut vacuum-vacuum diagrams with $r$-particle
cuts. (Recall that these are cuts that intercept $r$ propagators.) The
sum of these over all values of $r$ is therefore equal to $a/g^2$ by
definition.  Eq.~(\ref{eq:unitarity-cond}) therefore confirms that the
identity in eq.~(\ref{eq:a}) is precisely the unitarity condition.

\subsection{Moments of the distribution}
Moments of the distribution of probabilities
$P_n$ are easily computed with the generating function
\begin{equation}
F(x)\equiv \sum_{n=0}^{\infty}P_n \, e^{nx}\; ,
\label{eq:F-def}
\end{equation}
such that $\big<n^p\big>=F^{(p)}(0)$. 
Using eq.~(\ref{eq:Pn}), this generating function can be evaluated in
closed form, and one obtains
\begin{equation}
F(x) = e^{-a/g^2} \;
\exp\left[\frac{1}{g^2}\sum_{r=1}^{\infty}b_r e^{rx}\right]
=
\exp\left[\frac{1}{g^2}\sum_{r=1}^{\infty}b_r (e^{rx}-1)\right]\; .
\label{eq:Fx}
\end{equation}

The mean of the distribution of multiplicities, $\big<n\big>\equiv
\sum_n n P_n$, is
\begin{equation}
\big<n\big> = F^\prime(0)=\frac{1}{g^2}\sum_{r=1}^{\infty}r\, b_r \; .
\label{eq:avg-n}
\end{equation}
The average multiplicity is therefore given by the sum of all 
$r$-particle cuts through the connected vacuum-vacuum diagrams,
weighted by the number $r$ of particles on the cut. The
second derivative of $F(x)$ at $x=0$ is simply
\begin{equation}
F^{\prime\prime}(0)=\big<n^2\big>\; .
\end{equation}
The variance of the distribution, $\sigma\equiv
\big<n^2\big>-\big<n\big>^2$, can instead be obtained directly from  the second
derivative of
\begin{equation}
G(x)\equiv \ln F(x) = \frac{1}{g^2}\sum_{r=1}^{\infty}b_r (e^{rx}-1)
\end{equation}
at $x=0$. Thus, 
\begin{eqnarray}
\sigma
=G^{\prime\prime}(0)
=\frac{1}{g^2}\sum_{r=1}^{\infty}r^2 b_r \; .
\end{eqnarray}
More generally, the connected part of the moment of order $p$ reads
\begin{equation}
\big<n^p\big>_{\rm connected}=G^{(p)}(0)=
\frac{1}{g^2}\sum_{r=1}^{\infty}r^p b_r \; .
\end{equation}

One sees that if $b_r\not=0$ for at least one $r>1$, the variance and
the mean are not equal and the distribution is not a Poisson
distribution. The converse is true: it is trivial to check that
eq.~(\ref{eq:Pn}) is a Poisson distribution if $b_1\not=0$ and $b_r=0$
for $r>1$. Indeed, in this case, case, eq.~(\ref{eq:Pn}) would simply
be
\begin{equation}
P_n=e^{-b_1/g^2}\; \frac{1}{n!}\left(\frac{b_1}{g^2}\right)^n\; .
\label{eq:poisson}
\end{equation}
In a certain sense, eq.~(\ref{eq:Pn}) is approximately Poissonian,
since the term $(b_1/g^2)^n$ has the largest power of
$g^{-2}$ in $P_n$. However, even if this approximation is a good one for
the individual probabilities $P_n$, it leads to an error of relative
order unity for the value of the mean multiplicity; 
eq.~(\ref{eq:poisson}) gives $\big<n\big>=b_1/g^2$, while the
correct value is $\big<n\big>=g^{-2}\sum_r r\,b_r$.

\subsection{Formal derivation of eq.~(\ref{eq:Pn})}
\label{sec:Pn-1}
In section \ref{sec:Pn}, the probabilities $P_n$ for producing $n$
particles were expressed in terms of $b_r/g^2$. The latter are the
$r$-particle cuts in the sum of cut connected vacuum-vacuum
diagrams. We worked out the expressions for $P_n$ in terms of $b_r$
only for the cases $n=1,2,3$ and guessed a generalization of the
formula to an arbitrary $n$. We shall here derive more rigorously this
expression for $P_n$ and shall rewrite it formally to identify the
$b_r$ coefficients.

An important step in the forthcoming proof is to note that the
expectation value of the time ordered product of $n$ fields is
\begin{equation}
\big<0_{\rm out}\big| T\,\phi(x_1)\cdots\phi(x_n)\big|0_{\rm in}\big>
=
\frac{\delta}{i\delta j(x_1)}\cdots \frac{\delta}{i\delta j(x_n)}
e^{i{\cal V}[j]}\; .
\end{equation}
This formula is equivalent to the usual way of obtaining time-ordered
Green's functions from a generating functional. However, in our case,
the source $j$ is not a fictitious variable that one sets to zero at
the end but is instead the (physical) external source. We can also use
the standard LSZ reduction formula~\cite{ItzykZ1} to rewrite the
amplitude for producing $n$ particles as~\footnote{The wave-function
renormalization factors $Z$ correspond to self-energy corrections on
the cut propagators of the vacuum-vacuum diagrams.}
\begin{equation}
\big<\p_1\cdots \p_n{}_{\rm out}\big|0_{\rm in}\big>
=\frac{1}{Z^{n/2}}
\int 
\left[\prod_{i=1}^n d^4x_i \; e^{ip_i\cdot x_i}\;(\square_{x_i}+m^2)
\frac{\delta}{i\delta j(x_i)}\right]
\;
e^{i{\cal V}[j]}\; .
\label{eq:LSZ2}
\end{equation}
The probability for producing $n$-particles is given by the expression 
\begin{equation}
P_n = \frac{1}{n!}
\int \left[\prod_{i=1}^n \frac{d^3 \p_i}{(2\pi)^3 2E_i}\right] 
\left|\big<\p_1\cdots \p_n{}_{\rm out}\big|0_{\rm in}\big>\right|^2 \; ,
\label{eq:Pn-def}
\end{equation}
where $E_i \equiv \sqrt{{{\p_i}}^2 + m^2}$. Substituting the r.h.s. of
eq.~(\ref{eq:LSZ2}) in the above, and noting that
\begin{equation}
G_{+-}^0(x,y) = 
\int \frac{d^3 \p_i}{(2\pi)^3 2E_i} e^{ip\cdot (x-y)} \; ,
\label{eq:pm-propagator}
\end{equation}
is the Fourier transform of the expression in
eq.~(\ref{eq:Fourier-props}), we can write the probability $P_n$
directly as
\begin{eqnarray}
P_n=
\frac{1}{n!}
{\cal D}^n[j_+,j_-]
\;
\left.
e^{i{\cal V}[j_+]}\;e^{-i{\cal V}^*[j_-]}
\right|_{j_+=j_-=j}\; ,
\label{eq:Pn-1}
\end{eqnarray}
where ${\cal D}[j_+,j_-]$ is the operator
\begin{equation}
{\cal D}[j_+,j_-]\equiv \frac{1}{Z}
\int d^4x \,d^4y\; G_{+-}^0(x,y)\;
(\square_{x}\!+\!m^2)
(\square_{y}\!+\!m^2)
\frac{\delta}{\delta j_+(x)}\frac{\delta}{\delta j_-(y)}\; .
\label{eq:calD}
\end{equation}
The sources in the amplitude and the complex conjugate amplitude are
labeled as $j_+$ and $j_-$ respectively to ensure that the functional
derivatives act only on one of the two factors.

In eq.~(\ref{eq:Pn-1}), the mass $m$ in $G_{+-}^0$ and in
$\square+m^2$ is the physical pole mass of the particles under
consideration rather than their bare mass. The role of the prefactor
$Z^{-1}$ becomes more transparent if we write
\begin{equation}
{\cal D}=
\int d^4x\, d^4y \;
ZG_{+-}^0(x,y) \; \frac{\square_x+m^2}{Z}\;\frac{\square_y+m^2}{Z}
\frac{\delta}{\delta j_+(x)}\frac{\delta}{\delta j_-(y)}\; ,
\label{eq:D1}
\end{equation}
and recall that $Z$ is the residue of the pole of the renormalized
propagator. All the factors under the integral in eq.~(\ref{eq:D1})
are therefore renormalized propagators, or their inverse.

Using eq.~(\ref{eq:Pn-1}), we can write the generating function
$F(x)$, defined earlier in eq.~(\ref{eq:F-def}), as
\begin{equation}
F(x)=e^{e^x {\cal D}[j_+,j_-]} \;
\left.
e^{i{\cal V}[j_+]}\;e^{-i{\cal V}^*[j_-]}
\right|_{j_+=j_-=j}\; .
\label{eq:F2}
\end{equation}
The reader may note that the factor $\exp(i{\cal V}[j_+])$ represents
the sum of the usual Feynman vacuum-vacuum diagrams with source $j_+$.
Likewise, the factor $\exp(-i{\cal V}^*[j_-])$ is comprised of the
complex conjugate of vacuum-vacuum diagrams, with source $j_-$. One
thus begins to see a formal realization of the intuitive discussion of
the previous section in terms of cut diagrams. Alternately, the second
factor of eq.~(\ref{eq:F2}) can be interpreted as containing terms in
the Schwinger-Keldysh  expansion \cite{Schwi1,Keldy1} that have only
$+$ vertices, while the third factor contains terms that have only $-$
vertices. A brief introduction to the Schwinger-Keldysh formalism is
provided in appendix \ref{app:SK}.

We shall now further elaborate on the interpretation of $P_n$ (and
$F(x)$) as cut vacuum-vacuum Feynman diagrams as well as their
relation to the Schwinger-Keldysh framework. Consider the action of
the operator ${\cal D}[j_+,j_-]$ on the vacuum-vacuum factors in
eq.~(\ref{eq:F2}). The operator $Z^{-1}(\square_x+m^2)\delta/\delta
j_+(x)$ takes a diagram of $\exp(i{\cal V}[j_+])$, removes one of its
sources, and amputates the propagator to which this source was
attached.  (This procedure includes any self-energy decoration on this
propagator because of the factor $Z^{-1}$.)
$Z^{-1}(\square_y+m^2)\delta/\delta j_-(y)$ does the same on the
factor $\exp(-i{\cal V}^*[j_-])$. Finally, the factor $ZG_{+-}^0(x,y)$
is a (dressed) propagator that links the two diagrams that have been
opened by the functional derivatives with respect to the sources. When
$x=0$, one can convince oneself that the action of the operator
$\exp({\cal D}[j_+,j_-])$ is to build all the other vacuum-vacuum
diagrams of the Schwinger-Keldysh formalism, namely those that have at
least one $G_{-+}^0$ or $G_{+-}^0$,
\begin{equation}
e^{{\cal D}[j_+,j_-]} \;
e^{i{\cal V}[j_+]}\;e^{-i{\cal V}^*[j_-]}
=
e^{i{\cal V}_{_{SK}}[j_+,j_-]}\; .
\label{eq:SK-vacuum}
\end{equation}
Here $i{\cal V}_{_{SK}}[j_+,j_-]$ denotes the sum of all connected
vacuum-vacuum diagrams in the Schwinger-Keldysh formalism, with the
source $j_+$ on the upper branch of the contour and likewise, $j_-$ on
the lower branch.  When the magnitudes of the sources on the upper and
lower branches are the same, $j_+=j_-=j$, it is well known that this
sum of all vacuum-vacuum diagrams is equal to 1. (This statement is
proved in appendix \ref{app:SK}.) Therefore, eq.~(\ref{eq:F2})
satisfies $F(0)=1$, as unitarity dictates it should.

{\sl When $x\not=0$, the generating function $F(x)$ is the sum of all
vacuum-vacuum diagrams in the Schwinger-Keldysh formalism, with the
two ``off-diagonal" propagators substituted by}
\begin{eqnarray}
&& G_{-+}^0 \quad\to\quad e^x G_{-+}^0\; ,
\nonumber\\
&& G_{+-}^0 \quad\to\quad e^x G_{+-}^0\; ,
\label{eq:count}
\end{eqnarray}
{\sl and where the source is the same on the upper and lower branches
of the contour.}  Naturally, the logarithm of this generating function
is obtained by keeping only the connected vacuum-vacuum diagrams in
the sum.

Recall now that the building blocks of the Schwinger-Keldysh
perturbative expansion are the same as those obtained by the cutting
rules for calculating the imaginary parts of Feynman diagrams. We can
therefore identify each term in the expansion of $F(x)$ with a cut
vacuum-vacuum diagram. Moreover, one sees from eq.~(\ref{eq:count})
that the power of $e^x$ in a given term is equal to the number of
propagators $G_{-+}^0$ or $G_{+-}^0$ it contains. Therefore, the order
in $e^x$ equals the number of particles on the cut. Since $\ln(F(x))$
contains only the cut {\bf connected} vacuum-vacuum diagrams, and since
$b_r/g^2$ has been defined earlier as the sum of all the $r$-particle
cuts through connected vacuum-vacuum diagrams, we have just proved that
\begin{equation}
\ln(F(x))=-\frac{a}{g^2}+\sum_{r=1}^{\infty}e^{rx}\,\frac{b_r}{g^2}\; .
\end{equation}
The coefficients $b_r$ can thus be formally identified and computed in
principle.  One obtains the constant term in this formula, $-a/g^2$,
from the limit $x\to -\infty$ of eq.~(\ref{eq:F2}) and from
eq.~(\ref{eq:im-V}).

\section{Direct calculation of the average multiplicity}
\label{sec:keldysh}
In the previous section, we obtained a general expression for a
generating function which can be used, in principle, to compute all
moments of the multiplicity distribution. In this section, we will
concentrate on the first moment, the average multiplicity. We will
develop the Schwinger-Keldysh formalism, introduced previously in
section~\ref{sec:Pn-1} and in appendix \ref{app:SK}, to compute this
quantity both at leading and next-to-leading order. While the result
at leading order is well known, the next-to-leading order result is
new, and is the central result of this paper.

\subsection{General formula}
Although one could in principle compute the average multiplicity using
eq.~(\ref{eq:avg-n}), it would be extremely impractical because this
method requires one to evaluate separately all the $r$-particle
cuts through the vacuum-vacuum diagrams\footnote{This would be an
almost impossible task even at tree level.}.  It is much more
straightforward to obtain the multiplicity from the generating
function given in eq.~(\ref{eq:F2}). We obtain
\begin{eqnarray}
\big<n\big>=F^\prime(0)=
{\cal D}[j_+,j_-] \;\;
\left.
e^{i{\cal V}_{_{SK}}[j_+,j_-]}
\right|_{j_+=j_-=j}\; ,
\end{eqnarray}
where we have used eq.~(\ref{eq:SK-vacuum}). Writing out ${\cal
D}[j_+,j_-]$ explicitly using eq.~(\ref{eq:calD}), this expression can
be rewritten as
\begin{eqnarray}
&&\big<n\big>
=
\int d^4x d^4y\;
ZG_{+-}^0(x,y)\;
\frac{\square_x+m^2}{Z}\;\frac{\square_y+m^2}{Z}
\nonumber\\
&&\qquad\qquad\qquad\qquad
\times
\left[
\frac{\delta i{\cal V}_{_{SK}}}{\delta j_+(x)}
\frac{\delta i{\cal V}_{_{SK}}}{\delta j_-(y)}
+
\frac{\delta i{\cal V}_{_{SK}}}{\delta j_+(x)\delta j_-(y)}
\right]_{j_+=j_-=j}\; ..
\label{eq:n2}
\end{eqnarray}
Functional derivatives of the sum of connected
vacuum-vacuum diagrams are connected Green's functions with the 
number of external points equalling the number of derivatives. When we
differentiate $i{\cal V}_{_{SK}}$, 
\begin{equation}
\frac{\delta}{\delta j_{\epsilon_1}(x_1)}
\cdots
\frac{\delta}{\delta j_{\epsilon_n}(x_n)}
\;
i{\cal V}_{_{SK}}[j_+,j_-]
=\big<0_{\rm in}\big|P\,\phi^{(\epsilon_1)}(x_1)\cdots
\phi^{(\epsilon_n)}(x_n)\big|0_{\rm in}\big>_{\rm conn}\; ,
\label{eq:G-SK}
\end{equation}
the corresponding Green's functions are the path-ordered Green's
functions defined in eq.~(\ref{eq:SK-corr}) of appendix
\ref{app:SK}. More precisely, they are the connected components of
these Green's functions -- hence the subscript ``conn''. The operators
$Z^{-1}(\square+m^2)$ in eq.~(\ref{eq:n2}) amputate the external legs
of these Green's functions (along with the self-energy corrections
they may carry).

Denoting $\Gamma^{(\epsilon_1\cdots\epsilon_n)}(x_1,\cdots,x_n)$ as
the amputated version of the Green's function in eq.~(\ref{eq:G-SK}),
we finally express the average multiplicity as\footnote{If one is only
interested in the calculation of the multiplicity $\big<n\big>$, a
more elementary derivation is to start from
\begin{equation}
\big<n\big>=\int\frac{d^3\p}{(2\pi)^3 2 E_p}\;
\big<0_{\rm in}\big|a^\dagger_{\rm out}(\p)a_{\rm out}(\p)\big|0_{\rm in}\big>
\; .
\label{eq:nbar-a+a}
\end{equation}
A reduction formula for the correlator that appears under the integral
gives
\begin{eqnarray}
&&
\big<0_{\rm in}\big | a^\dagger_{\rm out}(\p)a_{\rm out}(\p) \big|0_{\rm in}\big>
=\frac{1}{Z}
\int d^4x d^4y\; e^{-ip\cdot x} e^{ip\cdot y}
\nonumber\\
&&\qquad\qquad\qquad\qquad\times
(\square_x+m^2)(\square_y+m^2)\;
\big<0_{\rm in}\big|\phi(x)\phi(y)\big|0_{\rm in}\big>\; .
\label{eq:reduc-nbar}
\end{eqnarray}
The expectation value on the right hand side, in which the fields are
not time-ordered, is just the propagator $G_{-+}(x,y)$ of the
Schwinger-Keldysh formalism. Splitting it into connected and
disconnected contributions, one can see that
eqs.~(\ref{eq:nbar-a+a})-(\ref{eq:reduc-nbar}) and (\ref{eq:n3}) are
equivalent.}
\begin{equation}
\big<n\big>
=
\int d^4x d^4y\;
ZG_{+-}^0(x,y)
\;\left[
\Gamma^{(+)}(x)\Gamma^{(-)}(y)+\Gamma^{(+-)}(x,y)
\right]_{j_+=j_-=j}\; .
\label{eq:n3}
\end{equation}
This formula is valid to all orders in the coupling constant and is a
key result of this paper as will become clearer later.
Diagrammatically, we can represent it as \setbox1\hbox to
4cm{\hfil\resizebox*{4cm}{!}{\includegraphics{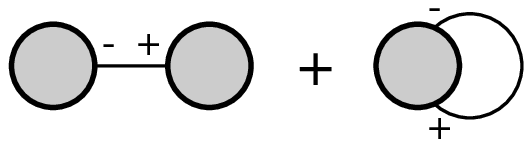}}}
\begin{equation}
\big<n\big>=\quad\raise -5mm\box1\quad ,
\label{eq:nbar1}
\end{equation}
where the shaded blobs represent the amputated Green's functions
$\Gamma$. In principle, one should amputate as well the self-energy
corrections attached to their external points and use the full $-+$
propagator. Alternatively, one can use the bare $-+$ propagator, and
keep the self-energy corrections attached to the external legs of the
shaded blobs.

Since the $\Gamma$'s are connected Green's functions, the second of
the two diagrams is at least a 1-loop diagram. Therefore only the
first diagram can contribute at leading order in the coupling. The
functions $\Gamma$ are obtained with the standard Schwinger-Keldysh
rules, described in the appendix \ref{app:SK}. Note also that, unlike
the expressions for the probabilities, there are no vacuum-vacuum
diagrams left in this formula. They have all disappeared thanks to the
fact that $\exp(i{\cal V}_{_{SK}}[j,j])=1$. This makes the evaluation
of the multiplicity much simpler than that of the probabilities $P_n$
even for $n=1$.

\subsection{Leading order result}
The multiplicity at leading order, namely O($g^{-2} (gj)^n)$, is
obtained from the left diagram in eq.~(\ref{eq:nbar1}) with each of
the two blobs evaluated at tree level. We have, \setbox1=\hbox to
1.8cm{\hfil\resizebox*{1.6cm}{!}{\includegraphics{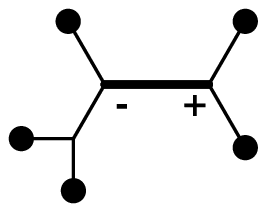}}}
\begin{equation}
\big<n\big>_{_{LO}}
=\sum_{+/-} \quad\raise -7mm\box1\quad,
\label{eq:phi-phi}
\end{equation}
where the sum is over all the tree diagrams on the left and on the
right of the propagator $G_{+-}^0$ (represented in boldface) as well
as a sum over the labels $+/-$ of the vertices whose type is not
written explicitly. At this order, the mass in $G_{+-}^0$ is simply
the bare mass, and $Z=1$.

Eq.~(\ref{eq:phi-phi}) can be expressed very simply
in terms of the {\bf retarded} solutions of the classical equations of
motion. The identities,
\begin{eqnarray}
&&
G^0_{++}-G^0_{+-}=G^0_{_R}\; ,
\nonumber\\
&&
G^0_{-+}-G^0_{--}=G^0_{_R}\; ,
\label{eq:r-trick}
\end{eqnarray}
where $G_{_R}^0$ is the free retarded propagator, will be essential in
demonstrating this result. Consider for instance, one of the tree
subdiagrams contained in eq.~(\ref{eq:phi-phi}). Summing over the $+$
or $-$ indices of the ``leaves'' (or ``blobs") of the tree, the outer
layer of propagators is transformed into retarded propagators, thanks
to eqs.~(\ref{eq:r-trick}). These then give factors of $\int d^4 y
G_R(\cdots,y) j(y)$ that are independent of the indices of the
vertices immediately below. This trick is applicable here because the
source $j(x)$ is identical on both branches of the closed time
path. By repeating this procedure recursively until one reaches the
``root'' of the tree, one turns all the propagators into retarded
propagators. As a consequence, one finally obtains the same tree
diagram (ending at point $x$ or $y$), where all the propagators are
now retarded propagators.

Now recall the perturbative
expansion of the retarded solution of the classical equation of
motion\footnote{To see that the various factors of $i$ present
in the Feynman rules of the Schwinger-Keldysh formalism agree with
this identification, simply multiply eq.~(\ref{eq:EOM}) by $i$, and
recall that we have defined the propagators as the inverse of
$i(\square+m^2)$ -- for instance, see eq.~(\ref{eq:Fourier-props}).}
\begin{equation}
(\square+m^2)\phi_c(x)+\frac{g}{2}\phi_c^2(x)=j(x)\; ,
\label{eq:EOM}
\end{equation}
with an initial condition at $x_0=-\infty$ such that the field and its
derivatives vanish
\begin{equation}
 \lim_{x^0\to -\infty}\phi_c(x)=0\; ,\quad 
\lim_{x^0\to -\infty}\partial^\mu\phi_c(x)=0\; .
\label{eq:init-cond}
\end{equation}
In that case as well, the solution can be expressed, identically,
recursively, in terms of the retarded Green's function.

The sum over all leaves up to the root of the tree on each side of the
cut in eq.~(\ref{eq:phi-phi}) can therefore be identified as
\setbox1=\hbox to
1.55cm{\hfil\resizebox*{1.55cm}{!}{\includegraphics{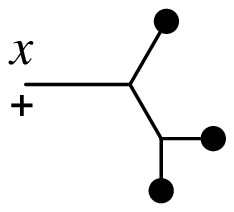}}}
\setbox2=\hbox to
1.55cm{\hfil\resizebox*{1.55cm}{!}{\includegraphics{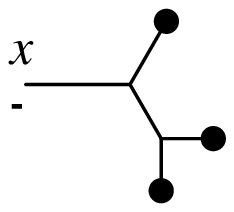}}}
\begin{equation}
\phi_c(x)=\sum_{+/-}\;\;\raise -8mm\box1 \quad
= \sum_{+/-}\;\;\raise -8mm\box2
\label{eq:phi_c}\quad.
\end{equation}
Note that in eq.~(\ref{eq:phi-phi}), we need to amputate these tree
diagrams, so it is $(\square+m^2)\phi_c(x)$ that enters in
eq.~(\ref{eq:phi-phi}) rather than $\phi_c(x)$ itself. More precisely,
we have
\begin{equation}
\big<n\big>_{_{LO}}=\int d^4x \,d^4y\; G^0_{+-}(x,y)\;
(\square_x+m^2)(\square_y+m^2)\,\phi_c(x)\phi_c(y)\; .
\end{equation}
Using the explicit momentum space form for $G_{+-}^0$ in
eq.~(\ref{eq:pm-propagator}),we can rewrite
\begin{equation}
\big<n\big>_{_{LO}}=\int\frac{d^3\p}{(2\pi)^32E_p}\;
\left|\int d^4x \;e^{ip\cdot x}\;(\square+m^2)\phi_c(x)\right|^2\; .
\label{eq:nbar}
\end{equation}
This leading order result is well known. In the CGC framework for
gluon production, the multiplicity was evaluated by computing the
spatial Fourier transform of the classical field $\phi_c$ at late
times~\cite{KovneMW1,KovneMW2,KrasnV1,KrasnV2,KrasnNV2,Lappi1}. To
connect their work to our eq.~(\ref{eq:nbar}), we use the identity
\begin{equation}
e^{ip\cdot x} (\partial_0^2+E_p^2)\phi_c(x) = 
\partial_0 \left(e^{ip\cdot x}\big[\partial_0-iE_p\big]\phi_c(x)\right) \, ,
\end{equation}
and perform an integration by parts of the r.h.s. of eq.~(\ref{eq:nbar})
(using the initial condition of eq.~(\ref{eq:init-cond})) to obtain
\begin{equation}
\int d^4x \;e^{ip\cdot x}\;(\square+m^2)\phi_c(x)
=
\lim_{x^0\to +\infty}
\int d^3\x \; e^{ip\cdot x}\;\left[\partial_{x_0}-iE_p\right]
\phi_c(x)\; .
\label{eq:AR}
\end{equation}
In practice, the r.h.s. of this equation is the preferred method for
evaluating the average multiplicity. One only needs to solve the
classical equation of motion with the initial conditions of
eq.~(\ref{eq:init-cond}) and perform a spatial Fourier transform of
the solution at the latest time.

\subsection{Next-to-leading order result}
\label{sec:NLO}
At next-to-leading order (NLO) in the coupling constant (${\cal
O}(g^0(gj)^n)$), one obtains two sorts of corrections to the leading
contribution of eq.~(\ref{eq:phi-phi}):
\begin{itemize}
\item[(i)] the right diagram in eq.~(\ref{eq:nbar1}) with the blob
evaluated at tree level \setbox1=\hbox to
2cm{\hfil\resizebox*{2cm}{!}{\includegraphics{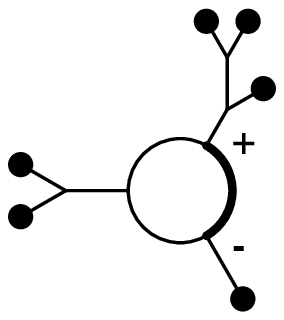}}}
\begin{equation}
\raise -7.5mm\box1
\label{eq:phi-phi-nlo2}
\end{equation}
\item[(ii)] 1-loop corrections to diagrams of the kind displayed in
eq.~(\ref{eq:phi-phi}). These come from the left diagram of
eq.~(\ref{eq:nbar1}), with one of the blobs evaluated at 1 loop, and
the other blob kept at tree level \setbox1=\hbox to
1.8cm{\hfil\resizebox*{1.8cm}{!}{\includegraphics{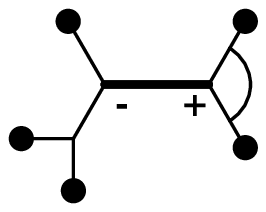}}}
\begin{equation}
\raise -8mm\box1
\label{eq:phi-phi-nlo1}
\end{equation}
\end{itemize}
In both cases, these contributions provide corrections of order
$g^0$ to the multiplicity $\big<n\big>$, to be compared to the
leading contributions of order $g^{-2}$.

That one obtains these two sets of contributions at NLO is also well
known. However, what is novel, and will be demonstrated below, is that
both these contributions can be computed entirely with retarded
boundary conditions and retarded propagators.  {\it This makes it
feasible to perform NLO computations (to all orders in the external
sources) by solving equations of motion for the classical and small
fluctuation fields with boundary conditions entirely specified at $t =
-\infty$.}

\subsubsection{Evaluation of eq.~(\ref{eq:phi-phi-nlo2})}
\label{sec:conn-trees}

We begin by discussing the diagrams displayed in
eq.~(\ref{eq:phi-phi-nlo2}). These topologies are well known since
they are the ones involved for the production of fermion-antifermion
pairs~\cite{BaltzGMP1}. (Indeed, for this process, they constitute the
lowest order contribution.) Evaluating the diagrams of
eq.~(\ref{eq:phi-phi-nlo2}) is equivalent to calculating the tree
level propagator $G_{+-}$ attached to an arbitrary number of tree
diagrams of the type depicted in eq.~(\ref{eq:phi_c}). Each of these
attachments, thanks to eq.~(\ref{eq:phi_c}), can be replaced by the
classical field $\phi_c$ itself. Thus \setbox1=\hbox to
2.8cm{\resizebox*{2.8cm}{!}{\includegraphics{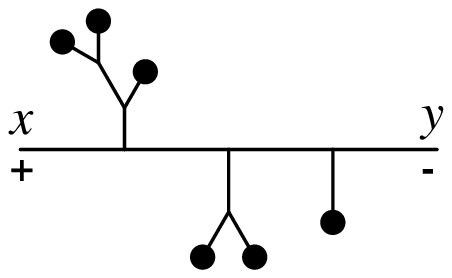}}}
\setbox2=\hbox to
2.8cm{\resizebox*{2.8cm}{!}{\includegraphics{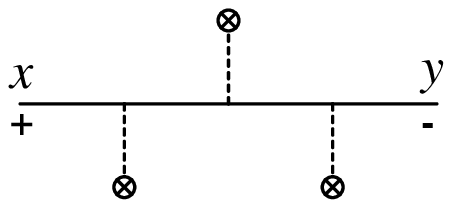}}}
\begin{equation}
\sum_{+/-}\quad\raise -7.5mm\box1 \quad=\quad 
\sum_{+/-}\quad\raise -7.5mm\box2\;\;\; ,
\end{equation}
where a dotted line terminating in a cross represents an insertion of
the classical field (eq. (\ref{eq:phi_c})). At the vertices
where $\phi_c$'s are inserted, the $+/-$ indices must still be summed 
over; a vertex of type $-$ simply corresponds to a sign change for the field
insertion. This is equivalent to computing the propagator
$G_{+-}(x,y)$ in the presence of the background field $\phi_c$. 
We shall mimic the method used in
\cite{BaltzGMP1} and write down a Lippmann-Schwinger equation whose
solution is $G_{+-}(x,y)$. For later reference, we shall write it 
for any component $G_{\epsilon\epsilon^\prime}(x,y)$ as 
\begin{equation}
G_{\epsilon\epsilon^\prime}(x,y)
=
G_{\epsilon\epsilon^\prime}^0(x,y)
-ig\sum_{\eta,\eta^\prime=\pm}\int d^4z \;
G_{\epsilon\eta}^0(x,z)
\;\phi_c(z){\bs\sigma}^3_{\eta\eta^\prime}\;
G_{\eta^\prime\epsilon^\prime}(z,y)\; ,
\label{eq:LS}
\end{equation}
where the resummed propagator has no superscript
0. ${\bs\sigma}^3=\mbox{diag}(1,-1)$ is the third Pauli matrix, and
its purpose in eq.~(\ref{eq:LS}) is to provide the correct minus sign
when the vertex at which the classical field is inserted is of type
$-$. 

The Lippman-Schwinger equation is solved by performing a simple
rotation\footnote{Such transformations were introduced in
\cite{AurenB1} and discussed more systematically in \cite{EijckKW1}.}
of the $+/-$ indices.  The details of this procedure can be followed
in appendix~\ref{app:LS}. The solution of the Lippmann-Schwinger
equation for the ``rotated'' propagator ${\bs G}$ reads
\begin{equation}
{\bs G}=
\begin{pmatrix}
0 & G_{_A} \cr
\cr
G_{_R} &
G_{_S}\cr
\end{pmatrix}
\; ,
\end{equation}
where the resummed retarded and advanced propagators are given by
\begin{equation}
G_{_R}=G_{_R}^0\sum_{n=0}^\infty\left[\phi_c G_{_R}^0\right]^n
\; ,\quad
 G_{_A}=G_{_A}^0\sum_{n=0}^\infty\left[\phi_c G_{_A}^0\right]^n\; ,
\end{equation}
and 
\begin{equation}
G_{_S}=G_{_R} \left(G_{_R}^0\right)^{-1} 
G_{_S}^0 
              \left(G_{_A}^0\right)^{-1}
G_{_A} \, .
\end{equation}
Here $G_{_R}^0$ and $G_{_A}^0$ are the free retarded and advanced
propagators respectively, and $G_{_S}^0 = G_{++}^0 + G_{--}^0 \equiv
2\pi \delta(p^2 - m^2)$.  The combination $G_{_R} G_{_R}^0{}^{-1}$ is
the full retarded propagator in the presence of the background field
$\phi_c$, amputated of its rightmost leg. The propagators
$G_{\epsilon\epsilon^\prime}$ of the Schwinger-Keldysh formalism are
obtained by performing the inverse of the rotations described in
appendix~\ref{app:LS} and the explicit expressions are given there.

To compute the next-to-leading order contribution to the multiplicity
$\big<n\big>$ from eq.~(\ref{eq:phi-phi-nlo2}), we need specifically
the propagator $G_{+-}$. The solution of the Lippman-Schwinger
equation gives
\begin{equation}
G_{+-}=G_{_R} G_{_R}^0{}^{-1} G_{+-}^0 G_{_A}^0{}^{-1}G_{_A}
\end{equation}
We are not done yet
because eq.~(\ref{eq:n3}) tells us that one has to amputate the
external legs of $G_{+-}$. This requirement is conveniently fulfilled by 
introducing ``retarded and advanced scattering T-matrices'',
defined by the relations 
\begin{eqnarray}
&&
G_{_R}\equiv G_{_R}^0 + G_{_R}^0 T_{_R} G_{_R}^0\; ,
\nonumber\\
&&
G_{_A}\equiv G_{_A}^0 + G_{_A}^0 T_{_A} G_{_A}^0\; .
\label{eq:TRA}
\end{eqnarray}
The contribution of eq.~(\ref{eq:phi-phi-nlo2}) to $\big<n\big>$ is
then given by
\begin{equation}
\big<n\big>_{_{NLO}}^{(1)}
\!=\!\int\!\!\frac{d^3\p}{(2\pi)^3 2E_p}
\!\int\! d^4x d^4y\, e^{-ip\cdot x} e^{ip\cdot y}
\!\!\int\! d^4u\, d^4v\, T_{_R}(x,u)G_{+-}^0(u-v)T_{_A}(v,y) \; .
\end{equation}
In momentum space, it reads more compactly as
\begin{equation}
\big<n\big>_{_{NLO}}^{(1)}
=\int\frac{d^3\p}{(2\pi)^3 2E_p}
\int \frac{d^3\q}{(2\pi)^3 2E_q}\;T_{_R}(p,-q)\,T_{_A}(-q,p)\; .
\end{equation}
Using the identity that $T_{_A}(-q,p)=[T_{_R}(p,-q)]^*$, we obtain finally, 
\begin{equation}
\big<n\big>_{_{NLO}}^{(1)}
=\int\frac{d^3\p}{(2\pi)^3 2E_p}
\int \frac{d^3\q}{(2\pi)^3 2E_q}\;\left|T_{_R}(p,-q)\right|^2\; .
\end{equation}
This formula is the equivalent for scalar bosons of eq.~(50) in
\cite{BaltzGMP1} and eq.~(94) in \cite{BlaizGV2}. 

The retarded $T$-matrix $T_R(p,-q)$ can be computed from the retarded
solution of the partial differential equation~\footnote{We remind the
reader that $g$ is a dimensionful quantity.},
\begin{equation}
\big(\square+m^2+g\phi_c(x)\big)\eta(x)=0\; ,
\label{eq:eom-lin}
\end{equation}
where $\eta$ is a small perturbation about the classical background
field $\phi_c$; the equation is therefore the linearized equation of
motion in this background.  Using Green's theorem\footnote{The
prefactor $i$ is due to the fact that the inverse of $\square+m^2$ is
$iG$ according to our conventions for the propagators.} (see appendix A
of \cite{BlaizGV1} for an extended discussion),
\begin{equation}
\eta(x)=i\int_{y_0=\mbox{const}} d^3{\vec\y}\;
G_{_R}(x,y)
\stackrel{\leftrightarrow}{\partial_{y_0}}\eta(y)\; ,
\end{equation}
as well as eq.~(\ref{eq:TRA}), one can prove that
\begin{equation}
T_{_R}(p,-q)=\lim_{x_0\to +\infty}
\int d^3\x \; e^{ip\cdot x}
\left[\partial_{x_0}-iE_p\right]
\eta(x)\; ,
\label{eq:FKT}
\end{equation}
where $\eta(x)$ is the retarded solution of eq.~(\ref{eq:eom-lin})
with initial condition $\eta(x)=e^{iq\cdot x}$ when $x_0\to -\infty$.
One should not confuse this equation with eq.~(\ref{eq:AR}), despite
the fact that they look similar. When using eq.~(\ref{eq:AR}) one must
solve the classical equation of motion in the presence of the source
$j$ in order to obtain $\phi_c$. In eq.~(\ref{eq:FKT}), one is instead
solving the equation of motion for a perturbation $\eta$ about the
classical solution $\phi_c$. The eigenvalues of this perturbation are
used to construct the small fluctuations propagator in the background
field. The analog of eq.~(\ref{eq:FKT}) for fermion production was
previously used in \cite{BlaizGV2} and \cite{GelisKL1,GelisKL2} to
evaluate the yield of quark-antiquark pairs in pA collisions and in
the initial stages of nucleus-nucleus collisions respectively.

\subsubsection{Evaluation of eq.~(\ref{eq:phi-phi-nlo1})}

We will now consider NLO contributions to the multiplicity from 1-loop
corrections such as those depicted in
eq.~(\ref{eq:phi-phi-nlo1}). Only the subdiagram that contains a
closed loop is new. Using the same trick as in
eq.~(\ref{eq:AR}), the contribution of eq.~(\ref{eq:phi-phi-nlo1}) to
$\big<n\big>_{_{NLO}}$ can be written as
\begin{eqnarray}
\big<n\big>_{_{NLO}}^{(2)}
&=&
\displaystyle{\int \frac{d^3\p}{(2\pi)^3 2E_p}}
\Big[
\lim_{x_0\to+\infty}\int d^3\x \; e^{ip\cdot x}
\big[\partial_0-iE_p\big]\phi_c(x)
\Big]\nonumber\\
&&\qquad\qquad\times
\Big[
\lim_{x_0\to+\infty}\int d^3\x \; e^{ip\cdot x}
\big[\partial_0-iE_p\big]\phi_{c,1}(x)
\Big]^*+\mbox{c.c.}\nonumber\\
&&
\end{eqnarray}
The novel quantity considered here is the 1-loop correction to
$\phi_c(x)$ -- we will represent this quantity as $\phi_{c,1}(x)$. We
will discuss below our strategy to compute this quantity by solving
partial differential equations with boundary conditions at $x^0
\rightarrow -\infty$.

All the diagrams contributing to $\phi_{c,1}(x)$
include a closed loop, to which may be attached an arbitrary number of
tree diagrams. All but one of these tree diagrams are terminated by
sources $j$; this last one has an external leg at the point $x$,
\setbox1=\hbox to
4.5cm{\resizebox*{4.5cm}{!}{\includegraphics{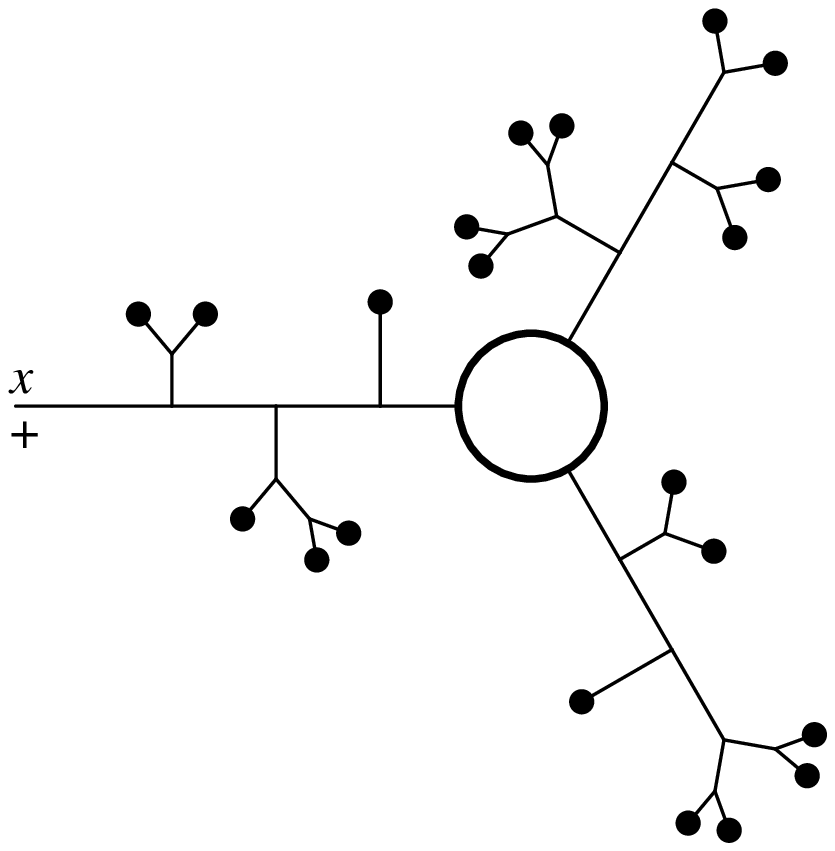}}}
\begin{equation}
\phi_{c,1}(x) = \sum_{+/-}\;\;\raise -23mm\box1
\end{equation}
The first simplification one can perform is to use
eq.~(\ref{eq:phi_c}) to collapse all the trees terminated by
sources into insertions of the classical field $\phi_c$. This gives
\setbox1=\hbox to
4cm{\resizebox*{4cm}{!}{\includegraphics{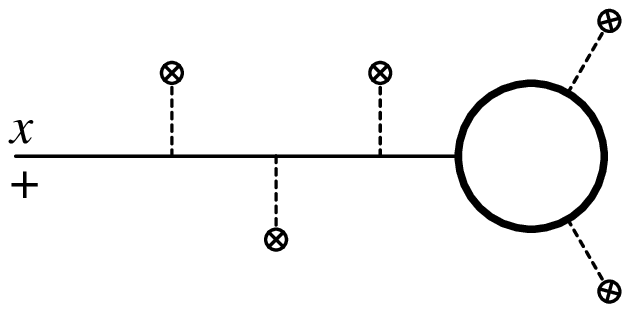}}}
\begin{equation}
\phi_{c,1}(x) = \sum_{+/-}\;\;\raise -9.5mm\box1
\label{eq:phi-c1-1}
\end{equation}
Again, at each remaining vertex or insertion of $\phi_c$, one must sum
over the type $+/-$. The number of insertions of $\phi_c$ can be
arbitrary, both on the loop and on the leg leading to the final point
$x$. Therefore, eq.~(\ref{eq:phi-c1-1}) can be written more explicitly
as
\begin{equation}
\phi_{c,1}(x) = -ig\sum_{\epsilon=\pm}\epsilon\,\int d^4y\;\;
G_{+\epsilon}(x,y)G_{\epsilon\epsilon}(y,y)\; ,
\label{eq:phi-c1-2}
\end{equation}
where all the propagators in this formula must be evaluated in the
presence of the background field $\phi_c$. 

We can rewrite $\phi_{c,1}$ as~\footnote{From eqs.~(\ref{eq:G1}) of
appendix~\ref{app:LS}, we see that the propagators $G_{++}(y,y)$ and
$G_{--}(y,y)$, which enter in eq.~(\ref{eq:phi-c1-2}), are both equal
to
\begin{equation}
G_{++}(y,y)=G_{--}(y,y)=
\frac{1}{2}
G_{_R} G_{_R}^0{}^{-1} G_{_S}^0 G_{_A}^0{}^{-1}G_{_A}\; .
\label{eq:G++2}
\end{equation}.}
\begin{eqnarray}
\phi_{c,1}(x) &=& -ig\,\int d^4y\;\;G_{++}(y,y)\sum_{\epsilon=\pm}\epsilon\;
G_{+\epsilon}(x,y)
\nonumber\\
&=&
-ig\int d^4y\;
G_{_R}(x,y)\;G_{++}(y,y)
\; ,
\label{eq:phi-c1-3}
\end{eqnarray}
where we have used eq.~(\ref{eq:r-trick}). This expression is
equivalent to the {\sl retarded} solution of the following partial
differential equation:
\begin{equation}
(\square+m^2+g\phi_c(x))\phi_{c,1}(x)=-g\,G_{++}(x,x)\; ,
\label{eq:ode2}
\end{equation}
with an initial condition such that $\phi_{c,1}$ and its derivatives
vanish at $x_0=-\infty$. The source term $G_{++}(x,x)$ in
eq.~(\ref{eq:ode2}) is given by eq.~(\ref{eq:G++2}),
\begin{eqnarray}
G_{++}(x,x)&=&\frac{1}{2}\int d^4u d^4v\;
\Big[G_{_R}G_{_R}^{0-1}\Big](x,u)\,
G_{_S}^0(u,v)\,\Big[G_{_A}^{0-1}G_{_A}\Big](v,x)
\nonumber\\
&=&\frac{1}{2}
\int \frac{d^4q}{(2\pi)^4}\;2\pi\delta(q^2-m^2)
\int d^4u \Big[G_{_R}G_{_R}^{0-1}\Big](x,u)\;e^{iq\cdot u}\nonumber\\
&&\qquad\qquad\qquad\qquad\quad\times
\int d^4v \Big[G_{_A}^{0-1}G_{_A}\Big](v,x)\;e^{-iq\cdot v}\; ,
\label{eq:G++3}
\end{eqnarray}
using
\begin{equation}
G_{_S}^0(u,v)=\int\frac{d^4q}{(2\pi)^4}\;
e^{iq\cdot(u-v)}
\;2\pi\delta(q^2-m^2)\; .
\end{equation}
{}From the r.h.s. of the relation,
\begin{equation}
\int d^4u \Big[G_{_R}G_{_R}^{0-1}\Big](x,u)\;e^{iq\cdot u} =
\lim_{u_0\to-\infty}i\int d^3\u \;
G_{_R}(x,u)\stackrel{\leftrightarrow}{\partial_{u_0}} e^{iq\cdot u}\equiv \eta(x)\,,
\end{equation}
we see that this quantity is precisely the retarded solution at
point $x$ of eq.~(\ref{eq:eom-lin}) with an initial condition
$\eta(u)=e^{iq\cdot u}$ at $u_0=-\infty$. Further, by noticing
that
\begin{equation}
\Big[G_{_A}^{0-1}G_{_A}\Big](v,x)
=
\Big[G_{_R}G_{_R}^{0-1}\Big](x,v)\; ,
\end{equation}
the factor involving $G_{_A}^{0-1}G_{_A}$ in eq.~(\ref{eq:G++3}) is
obtained in a similar way, except that the initial condition must be
$e^{-iq\cdot v}$.

Following this discussion, we can now outline here an algorithm for
obtaining the quantity $\phi_{c,1}(x)$ needed in the calculation of
$\big<n\big>_{_{NLO}}^{(2)}$:
\begin{itemize}
\item[(i)] For an arbitrary on-shell momentum $q$, find the solutions
$\eta_q^{(+)}(x)$ and $\eta_q^{(-)}(x)$ of eq.~(\ref{eq:eom-lin})
whose initial conditions at $x_0\to-\infty$ are respectively
$\eta_q^{(+)}(x)=e^{iq\cdot x}$ and $\eta_q^{(-)}(x)=e^{-iq\cdot x}$
\item[(ii)] Calculate $G_{++}(x,x)$ as
\begin{equation}
G_{++}(x,x)=\frac{1}{2}\int\frac{d^4q}{(2\pi)^4}\;2\pi\delta(q^2-m^2)\;
\eta_q^{(+)}(x)\eta_q^{(-)}(x)\; .
\label{eq:G++4}
\end{equation}
\item[(iii)] Solve eq.~(\ref{eq:ode2}) with a vanishing initial condition at
$x_0=-\infty$
\end{itemize}

For completeness, we note that $G_{++}(x,x)$ contains ultraviolet
divergences. They appear as a divergence in the integration over the
momentum $q$ in eq.~(\ref{eq:G++4}). By going back to the diagrams
that are contained in $G_{++}(x,x)$, it is easy to see that these
divergences are the usual 1-loop ultraviolet divergences of the
$\phi^3$ field theory in the vacuum, namely, the tadpole and the
self-energy divergences\footnote{Of course, the scalar toy model with
a $\phi^3$ self-interaction that we are considering in this paper is
somewhat peculiar in the ultraviolet sector, since this theory is
super-renormalizable in four dimensions. But the procedure outlined
here for removing the divergences would be valid in QCD as well.}.
These divergences correspond to the two terms of
eq.~(\ref{eq:phi-c1-1}) in which the loop is dressed by zero or one
insertions of the classical field $\phi_c$ (the terms with two
insertions of $\phi_c$ or more on the loop involve the 1-loop
corrections to the $n$-point functions for $n\ge 3$, which are finite
in four dimensions). From this observation, it is clear that one must
subtract the following quantity from $G_{++}(x,x)$
\begin{equation}
\delta G_{++}(x,x)
\equiv
\delta_1+(\delta_{_Z}\square+\delta_{m^2})\phi_c(x)\; ,
\label{eq:dG++}
\end{equation}
where $\delta_1$ is the counterterm that makes the tadpole finite, and
$\delta_{_Z}$ and $\delta_{m^2}$ the two counterterms that make the
self-energy finite. One possible approach is to evaluate both
eq.~(\ref{eq:G++4}) and the counterterms with the same momentum cutoff
$\Lambda$, and to subtract eq.~(\ref{eq:dG++}) from
eq.~(\ref{eq:G++4}) before letting $\Lambda$ go to infinity.

\section{AGK cancellations}
\label{sec:agk}

We will discuss here a set of combinatorial rules that relate the
discussions of the $n$-particle probabilities in section
\ref{sec:distribution} and the average multiplicity in the previous
section. We find that these combinatorial relations are identical to
the Abramovsky--Gribov--Kancheli (AGK) cancellations first
discussed~\cite{AbramGK1} in the context of reggeon field theory. This
enables us to map out explicitly, in the context of multi-particle
production, the terminology of field theories with external sources
(such as the Color Glass Condensate) with that of reggeon field
theory. Our discussion is in fact more general than the original AGK
discussion, since it is not limited to diagrams that have an
interpretation in terms of reggeons alone.

Before we proceed, we shall provide here a very brief sketch of the
extensive literature on the subject of the AGK cutting rules.  A
review of early work can be found in ref.~\cite{Weis1}.  Applications
of AGK rules to reggeon field theory models of particle production in
proton-nucleus and nucleus-nucleus collisions were discussed at length
by Koplik and Mueller~\cite{KopliM1}. More recently, several authors
considered the application of these rules in perturbative QCD. For a
comprehensive recent discussion, see Ref.~\cite{BarteR1,BarteSV1} and
references therein. We note that AGK rules have also been considered
previously in the Mueller dipole picture~\cite{Muell1} of
saturation/CGC~\cite{KovchL1,KovchT1}. A detailed discussion and
additional references can be found in \cite{JalilK1}.

\subsection{AGK-like identities}
\label{sec:agk-ids}
The AGK relations are a set of identities that relate the average
multiplicity of particles produced in hadronic collisions to the
multiplicity of a single cut reggeon. Thanks to these identities, all
the other contributions to the average multiplicity involving more
than one reggeon exchange (where the reggeons that are not cut
correspond to absorptive corrections) cancel in the calculation of the
average multiplicity.

Even though our discussion is based on a completely different theory,
one can obtain a set of identities that play the same role as the
original AGK identities\footnote{In this sense, one can understand
these identities to be a very general consequence of field theories
with strong external sources.}. To see these, consider $P_{n,m}$, the
term that has a factor $1/g^{2m}$ in the expression eq.~(\ref{eq:Pn})
for $P_n$,
\begin{equation}
P_{n,m}=\frac{1}{g^{2m}}\sum_{p+q=m}\frac{(-a)^q}{q!}\frac{1}{p!}
\sum_{\alpha_1+\cdots+\alpha_p=n}b_{\alpha_1}\cdots b_{\alpha_p}\; ..
\label{eq:Pnm}
\end{equation}
Obviously one has
\begin{equation}
P_n=\sum_{m=0}^{\infty} P_{n,m}\; .
\end{equation}
Here $m$ is the number of disconnected subdiagrams
contributing to $P_n$; each disconnected diagram starts at order
$g^{-2}$.  In eq.~(\ref{eq:Pnm}), $q$ is the number of 
disconnected subdiagrams that are not cut -- they do not contribute to
the number $n$ of produced particles, and correspond to the
``absorptive corrections'' of \cite{AbramGK1}. Likewise, $p$ is the number
of subdiagrams that are cut -- they contribute to the number
of produced particles. In the model of \cite{AbramGK1}, $m$ would be
the number of reggeons in the diagram.  

There is however one important
difference between our considerations and those of \cite{AbramGK1}. In
our paper, $n$ is the number of produced {\bf particles}, while in
\cite{AbramGK1} $n$ is the number of {\bf cut reggeons},
regardless of how many particles are produced in each of them. Therefore, 
in our case, $n$ can be larger than $m$  (because each
cut subdiagram can produce many particles), while $n$ is always smaller
or equal to $m$ in \cite{AbramGK1}.

We shall now introduce the term of order $1/g^{2m}$, $F_m(x)$, in the
generating function $F(x)$ introduced in section 3. It is defined as
\begin{equation}
F_m(x)
\equiv \sum_{n=0}^{\infty}  P_{n,m} \;e^{nx}
= \frac{1}{m!}\frac{1}{g^{2m}}\left(\sum_{r=0}^\infty b_r e^{rx} -a\right)^m\; .
\end{equation}
This formula can be obtained either by using eq.~(\ref{eq:Pnm}) and
performing the sum explicitly, or by expanding eq.~(\ref{eq:Fx}) to
the appropriate order $1/g^{2m}$. Evaluating the successive
derivatives of $F_m(x)$ at $x=0$, we can easily prove the identities
\begin{eqnarray}
&&
\mbox{if\ }m\ge2\;,\qquad \sum_{n=1}^\infty n P_{n,m} =0\; ,
\nonumber\\
&&
\mbox{if\ }m\ge3\;,\qquad \sum_{n=2}^\infty n(n-1) P_{n,m} =0\; ,
\nonumber\\
&&
\cdots
\end{eqnarray}
To prove these formulas, we used the unitarity relation, $a=\sum_r
b_r$.  The physical interpretation of these formulas is quite
transparent.  They tell us that Feynman diagrams that have two or more
disconnected subdiagrams do not affect the mean multiplicity, that
Feynman diagrams with three or more disconnected subdiagrams do not
affect the variance, and so on... These identities are the
generalization to our situation of the eqs.~(24) of \cite{AbramGK1}.

Note that the sum over $n$ must be extended to infinity 
due to the above mentioned fact that we are counting particles
instead of cut reggeons.

\subsection{Closer connection with reggeon theory}
\label{sec:reggeon}
One can in fact make more detailed connections with the discussion of
ref. \cite{AbramGK1}, and in a sense, generalize the results
there. Recall that the latter is formulated in the language of
reggeons so the connections to our framework are not obvious. We
remind the reader that in eq.~(\ref{eq:Pn}), for $P_n$, the index $p$
represents the number of cut subdiagrams contributing to the
production of the $n$ particles. Therefore, one can further divide
$P_n$ and write the probability for having an event in which $n$
particles are produced in $p$ cut subdiagrams\footnote{This new
quantity ${P}^{(c)}_{n,p}$ should not be confused with $P_{n,m}$. The
latter is the probability for producing $n$ particles in $m$
subdiagrams, irrespective of whether cut or uncut.  Defining
$P_{n,p,q}$ as the probability of producing $n$ particles with $p$ cut
subdiagrams and $q$ uncut subdiagrams, the former quantities are
respectively $P_{n,m}=\sum_{p+q=m}P_{n,p,q}$ and
${P}^{(c)}_{n,p}=\sum_{q=0}^{+\infty}P_{n,p,q}$.}
\begin{equation}
{P}^{(c)}_{n,p}=e^{-a/g^2}\;\frac{1}{p!}
\sum_{\alpha_1+\cdots+\alpha_p=n}
\frac{b_{\alpha_1}\cdots b_{\alpha_p}}{g^{2p}}
\; ,
\label{eq:regg-Pnp}
\end{equation}
which of course is related to $P_n$ by
\begin{equation}
P_n=\sum_{p=0}^n {P}^{(c)}_{n,p}\; .
\end{equation}
The unspecified ``subdiagrams'' in our discussion are a generalization
of the ``reggeons'' of \cite{AbramGK1}. Since the discussion in
\cite{AbramGK1} was at the level of the cut reggeons rather than at
the level of the individual particles, we may obtain a closer
correspondence to the results in ~\cite{AbramGK1} by integrating out
the index $n$ representing the number of produced particles. Summing
over $n$ in eq.~(\ref{eq:regg-Pnp}) gives the probability of having
$p$ cut subdiagrams,
\begin{equation}
{\cal R}_p\equiv\sum_{n=p}^{+\infty}{P}^{(c)}_{n,p}=
\frac{1}{p!}\left(\frac{a}{g^2}\right)^p\;e^{-a/g^2}\; .
\label{eq:regg-P}
\end{equation}
Clearly, the multiplicity of cut subdiagrams has a Poissonian
distribution.  This is not a surprise because they are disconnected
from one another\footnote{In \cite{AbramGK1}, this result was achieved
thanks to an assumption about the impact factors, while in our model
it follows naturally because particle production arises only through
the coupling to an external source.}. Moreover, the average number of
cut subdiagrams is equal to
\begin{equation}
\big<n_{\rm cut}\big>\equiv\sum_{p=0}^{+\infty}p{\cal R}_p=\frac{a}{g^2}\; .
\end{equation}
One may define the average number of particles produced in diagrams
with $p$ cut subdiagrams as\footnote{This quantity must be
defined as a conditional probability. Indeed, the denominator is
necessary in this definition in order to take into account the fact
that events with different numbers of cut subdiagrams can occur with different
probabilities.}
\begin{equation}
\big<n\big>_p\equiv \frac{\sum_{n=p}^{+\infty}n{\cal
P}^{(c)}_{n,p}}{\sum_{n=p}^{+\infty}{\cal P}^{(c)}_{n,p}}
=p\,\frac{\sum_{r=1}^{+\infty}rb_r}{a} \; .
\end{equation}
This result implies that 
\begin{equation}
\big<n\big>_p=p\;\big<n\big>_1\; .
\label{eq:nfact}
\end{equation}
In other words, a diagram with $p$ cut subdiagrams produces on average
$p$ times the number of particles produced by a diagram with a single
cut subdiagram. Further, one can deduce immediately that the average
number of produced particles, $\big<n\big>=g^{-2}\sum_r r b_r$, is the
average number of cut subdiagrams multiplied by the average number of
particles produced in one cut subdiagram,
\begin{equation}
\big<n\big>=\big<n_{\rm cut}\big>\;\big<n\big>_1\; .
\end{equation}
This property was assumed in ref.~\cite{AbramGK1}; they therefore
focused only on properties of the distribution of cut reggeons. In our
approach, this property arises naturally, regardless of the nature of
the subdiagrams we consider. (This result is therefore 
more general than the ladder diagrams that correspond to
reggeons.)

In order to exactly reproduce the identities in \cite{AbramGK1}, we
introduce the probability ${\cal R}_{p,m}$ of having $p$ cut
subdiagrams for a total number $m$ of subdiagrams ($m-p$ of which are
therefore uncut).  Expanding the exponential in eq.~(\ref{eq:regg-P})
to order $m-p$, one obtains
\begin{equation}
{\cal R}_{p,m}=\frac{1}{(m-p)!}\frac{1}{p!}\;
\left(-\frac{a}{g^2}\right)^{m-p}\left(\frac{a}{g^2}\right)^p\; .
\end{equation}
This distribution of
probabilities then satisfies the relations
\begin{eqnarray}
&&
\mbox{if\ }m\ge2\;,\qquad \sum_{p=1}^m p {\cal R}_{p,m} =0\; ,
\nonumber\\
&&
\mbox{if\ }m\ge3\;,\qquad \sum_{p=2}^m p(p-1) {\cal R}_{p,m} =0\; ,
\nonumber\\
&&
\cdots
\label{eq:agk-ident}
\end{eqnarray}
This set of identities is strictly equivalent to the eqs.~(24) of
\cite{AbramGK1}. For instance, the first relation means that diagrams
with two or more subdiagrams cancel in the calculation of the
multiplicity. These relations are therefore a straightforward
consequence of the fact that the distribution of the numbers of cut
subdiagrams is a Poisson distribution. They do not depend at all on
whether these subdiagrams are ``reggeons'' or not. In appendix
\ref{app:inv-agk}, we prove the reverse result: the only distribution
${\cal R}_p$ for which eqs.~(\ref{eq:agk-ident}) hold is a Poisson
distribution.

It is also instructive to note that these identities obeyed by the
distribution of cut/uncut disconnected subdiagrams have been obtained
by ``integrating out'' the number $n$ of produced particles. By doing
so, one goes from the distribution (\ref{eq:regg-Pnp}) -- which still
depends on all the quantities $b_1, b_2,b_3\cdots$ -- to
(\ref{eq:regg-P}), in which all the dynamics is summarized in one
number $a=\sum_r b_r$. Integrating out the number of particles has the
virtue of hiding most of the details of the field theory under
consideration. This is in fact the reason why the AGK identities
have a range of validity which is much wider than the reggeons models
for which they had been derived originally. 

However, this robustness has a cost: by integrating out the number of
particles, one has lost a lot of information about the details of the
theory, which means that certain questions can no longer be
addressed. For instance, in eq.~(\ref{eq:nfact}), the factor
$\big<n\big>_1$ -- the average number of particles produced in one cut
subdiagram -- can only be calculated by going back to the microscopic
dynamics of the theory under consideration.

\subsection{Additional cancellations in $\big<n\big>$}
In the original approach of Abramovsky, Gribov and Kancheli, one would
estimate the particle multiplicity predicted by the reggeon model as
follows:
\begin{itemize}
\item[(i)] compute the diagram with one cut reggeon exchange in order to
obtain the average number of cut reggeons, $\big<n_{\rm cut}\big>$
\item[(ii)] calculate the average number of particles in one cut reggeon
$\big<n\big>_1$. This second step in general requires a more
microscopic description of what one means by ``reggeons''
\end{itemize}

The advantage of our approach is of course that it is entirely
formulated as a field theory in which the elementary objects are
particles. This makes it possible to compute these two
quantities from first principles. Indeed, expressions
for these in terms of cuts through vacuum-vacuum Feynman diagrams are 
\begin{equation}
\big<n_{\rm cut}\big>=\frac{a}{g^2}\; ,\quad
\big<n\big>_1=\frac{g^{-2}\sum_r rb_r}{a g^{-2}}\; .
\label{eq:beyond-agk}
\end{equation}
As one can see, in our microscopic description, there is a further
cancellation in their product, between $\big<n_{\rm cut}\big>$ and the
denominator of $\big<n\big>_1$. This result could be anticipated because
we already know that the average multiplicity is equal to
$g^{-2}\sum_r rb_r$.

Moreover, by comparing eqs.~(\ref{eq:avg-n}) and (\ref{eq:phi-phi}),
we see that we have the following identity at leading order, 
\begin{equation}
\setbox1=\hbox to 1.8cm{\hfil\resizebox*{1.8cm}{!}{\includegraphics{fig07.ps}}}
\frac{1}{g^2}\sum_{r=1}^{\infty}\left.r b_r\right|_{_{LO}}
=
\sum\;\;\raise -8mm\box1
\label{eq:agk-rhs2}
\end{equation}
In the r.h.s. of this relation, the types of the endpoints of the
propagator represented in boldface must be held fixed, and one must
sum over the topologies and over the $+/-$ indices for the trees on
both sides of this propagator. The diagrammatic expansion of the left
hand side of eq.~(\ref{eq:agk-rhs2}) is of course obtained from
eqs.~(\ref{eq:b1}), (\ref{eq:b2}), (\ref{eq:b3}), etc... We also know
from eq.~(\ref{eq:phi_c}) that the tree diagrams that appear in the
r.h.s. are nothing but the retarded solution of the classical equation
of motion. In other words, eq.~(\ref{eq:agk-rhs2}) tells us that the
sum of all cuts through ({\bf time-ordered}) tree connected
vacuum-vacuum diagrams, weighted by the particle multiplicity on the
cut, can be cast into a much simpler expression in terms of {\bf
  retarded} tree diagrams. This means that additional cancellations
must occur in the left hand side of eq.~(\ref{eq:agk-rhs2}) in order
to have this dramatic simplification. These cancellations are
discussed explicitly in appendix~\ref{app:cancellations}.

\subsection{Higher moments}
\label{sec:moments}
The cancellations discussed in the previous subsection, which are in
fact responsible for the identity of eq.~(\ref{eq:agk-rhs2}), are
crucial in order to be able to calculate the particle multiplicity in
a simple way. We have seen explicitly that one can calculate the
multiplicity at leading order directly from the classical field
$\phi_c$, and in principle at higher orders as well. Of course, the
NLO calculation has the complications of a typical 1-loop calculation; 
in particular one has to treat the ultraviolet divergences.

It turns out that higher moments of the multiplicity distribution are
also calculable from $\phi_c$, by formulas that are tractable even
though they are more complicated than in the case of the multiplicity.
Let us consider the case of the variance as an illustration. It is
obtained as the second derivative at $x=0$ of the log of the
generating function, $\ln(F(x))$. Therefore, it is given by
\begin{equation}
\sigma\equiv\big<n^2\big>-\big<n\big>^2
=
\left[
{\cal D}[j_+,j_-]
+
{\cal D}^2[j_+,j_-]
\right]\;\;\left. e^{i{\cal V}_{_{SK}}[j_+,j_-]}
\right|_{{j_+=j_-=j}\atop{\rm connected}}\; .
\end{equation}
Performing the functional derivatives and dropping the disconnected
terms that show up gives
\begin{eqnarray}
&&\sigma
=\big<n\big>
+
\int d^4x d^4y d^4u d^4v \; G_{+-}^0(x,y)G_{+-}^0(u,v)\;
\Big[
 \Gamma^{(+-+-)}(x,y,u,v)
\nonumber\\
&&\qquad
+\Gamma^{(-++)}(y,u,v)\Gamma^{(+)}(x)
+\Gamma^{(++-)}(x,u,v)\Gamma^{(-)}(y)
+(x\leftrightarrow u,y\leftrightarrow v)
\nonumber\\
&&\qquad
+\Gamma^{(++)}(x,u)\Gamma^{(--)}(y,v)
+\Gamma^{(+-)}(x,v)\Gamma^{(-+)}(y,u)
\nonumber\\
&&\qquad
+\Gamma^{(+-)}(x,v)\Gamma^{(-)}(y)\Gamma^{(+)}(u)
+\Gamma^{(-+)}(y,u)\Gamma^{(+)}(x)\Gamma^{(-)}(v)
\nonumber\\
&&\qquad
+\Gamma^{(++)}(x,u)\Gamma^{(-)}(y)\Gamma^{(-)}(v)
+\Gamma^{(--)}(y,v)\Gamma^{(+)}(x)\Gamma^{(+)}(u)
\Big]\; , 
\end{eqnarray}
where the $\Gamma$'s are amputated Green's functions in the
Schwinger-Keldysh formalism.  Diagrammatically, the variance can be
expressed as \setbox1\hbox to
10.5cm{\hfil\resizebox*{10.5cm}{!}{\includegraphics{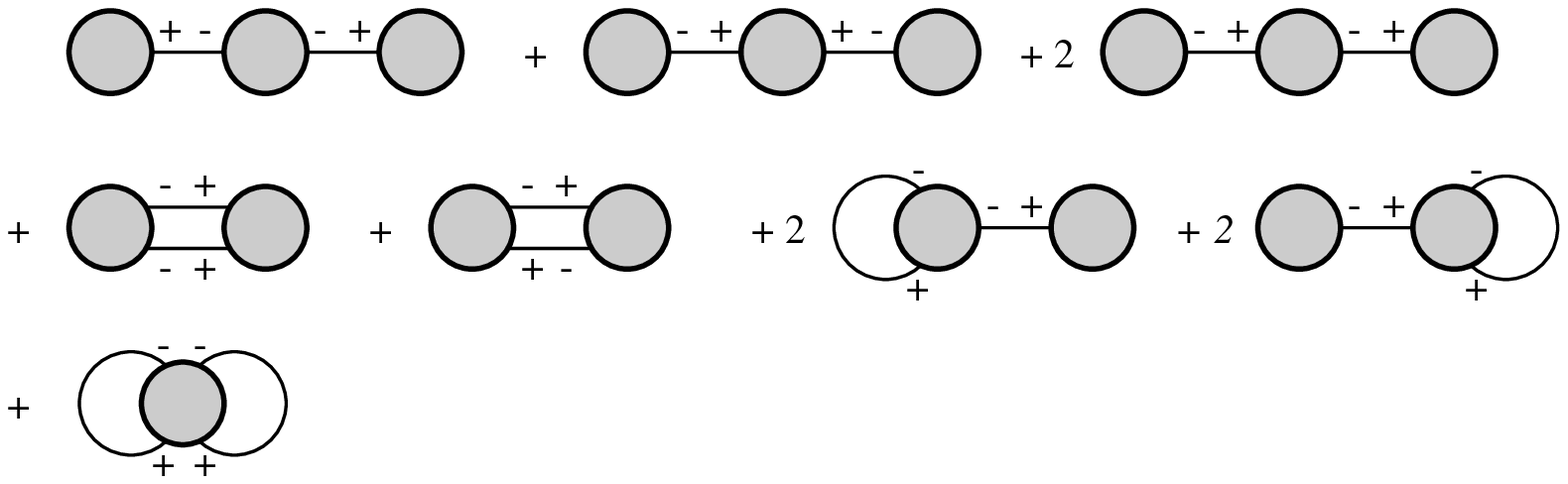}}}
\begin{equation}
\sigma=\big<n\big>
\;+\raise -28mm\box1
\end{equation}
All the diagrams represented in this equation are deviations from a
Poisson distribution (for which we would simply have
$\sigma=\big<n\big>$). The diagrams of the first line start at tree
level, while those of the second and third lines start respectively at
1 and 2 loops. Similar formulas can be derived for even higher
moments, but they become increasingly complicated as the order of the
moment increases, even at tree level.

\section{Summary and Outlook}

In this paper, we studied particle production in a model $\phi^3$
scalar field theory coupled to a strong ($j\sim 1/g$) time dependent
external source. This theory is a toy model of the Color Glass
Condensate formalism for particle production in high energy hadronic
collisions in QCD. We shall summarize here some of the novel results of this 
paper.

\begin{itemize}

\item We obtained a general formula (eq.~(\ref{eq:Pn})) for the
probability $P_n$ to produce $n$ particles in a field theory with time
dependent external sources. (A formal expression of this formula is
given in eqs.~(\ref{eq:Pn-1}) and (\ref{eq:calD}).)

\item We derived the corresponding generation function for moments of
the multiplicity distribution. We showed that the generating function
(eq.~(\ref{eq:F2})) can be related to the sum of all vacuum-vacuum
diagrams in the Sch\-win\-ger-Keldysh formalism with off-diagonal
propagators given by eq.~(\ref{eq:count}).

\item We obtained a general formula (eq.~(\ref{eq:n3})) for the average
particle multiplicity in terms of the amputated Schwinger--Keldysh
Green's functions. While the result to leading order is well known,
the result at next-to-leading order (NLO) is new and remarkable. It is
not surprising of course that the average multiplicity at NLO can be
expressed in terms of small fluctuation fields. What is non-trivial
and remarkable is that these fields can be computed by solving small
fluctuation equations of motion with {\it retarded} propagators and
initial conditions specified at $x^0\rightarrow -\infty$
(eqs.~(\ref{eq:eom-lin}), (\ref{eq:ode2}) and (\ref{eq:G++4})).

\item We outlined an algorithm to compute these NLO
contributions. Such an algorithm has been implemented previously for
fermion pair production in the Color Glass Condensate, where the
diagram of eq.~(\ref{eq:phi-phi-nlo2}) is the leading contribution. It
has not been implemented for the diagram of
eq.~(\ref{eq:phi-phi-nlo1}).

\item We observed that the Abramovsky-Gribov-Kancheli cancellations,
originally formulated in terms of relations among reggeon diagrams,
can be mapped into the field theory language of cut vacuum-vacuum
diagrams.  Disconnected vacuum-vacuum subgraphs can be identified as
reggeons and corresponding cut sub-graphs as cut reggeons.

\item We showed that an immediate consequence of this mapping is that
AGK cancellations in reggeon field theory hold if and only if the
distribution of cut reggeons is Poissonian. This follows because two
disconnected cut sub-diagrams do not interact by definition! The
conjecture is proved explicitly in appendix~\ref{app:inv-agk}. We
observed that there are additional cancellations (as represented by
eq.~(\ref{eq:agk-rhs2}) ) that are not visible at the level of the AGK
cancellations.

\item We derived a general formula for the second moment (the
variance) of the multiplicity distribution. This formula can be
expressed in terms of the average multiplicity plus ``non-Poissonian"
terms. These are nonzero even at tree level in a field theory with
strong time dependent external sources.

\end{itemize}

We shall now discuss some of the ramifications of our results, in
particular for high energy QCD.  Our studies were indeed motivated by
their possible relevance to hadronic collisions at very high
energies. In particular, in the Color Glass Condensate approach,
hadronic collisions at very high energies are described by an
effective field theory where the degrees of freedom are wee parton
fields coupled to strong time dependent external valence sources. A
key feature of the CGC approach is that weak coupling techniques are
applicable. The techniques developed here can therefore be applied to
study particle production\footnote{However, one should keep in mind
the fact that transition amplitudes in a slowly varying background
might be difficult to define rigorously \cite{Dietr3}.}. Of particular
interest is particle production in the earliest stages of heavy ion
collisions, and the possibility that one can understand the
thermalization of a quark gluon plasma from first principles. As is
immediately clear from the formalism developed here, the solution of
such problems is intrinsically non--perturbative (even in weak
coupling) requiring the summation of infinite series of Feynman
graphs. Numerical algorithms performing these summations are therefore
essential.
 
Indeed, first steps in carrying out the program outlined here have
already been performed. Gluon
production~\cite{KrasnV4,KrasnV1,KrasnV2,KrasnNV1,KrasnNV2,Lappi1} and
quark pair production~\cite{GelisKL1,GelisKL2} have been computed
numerically previously to leading order in the coupling and to all
orders in the source density. There are strong hints that a deeper
understanding of thermalization requires an analysis incorporating
energy loss effects~\cite{GyulaVWZ1,BaierSZ1,KovneW2}.  These may
include the scattering contributions of the bottom-up
scenario~\cite{BaierMSS2,MuellSW1}, or equivalently the energy loss
induced by collective effects~\cite{ArnolLMY1,RebhaRS1,Mrowc1}.  In
the CGC framework, these effects can be shown to appear at
next-to-leading order in the coupling. Qualitative
studies~\cite{KharzT2,RomatV1} of the NLO effects suggest that these
may be very important in driving the system towards equilibrium. Our
study here is a first step towards a quantitative formulation of this
problem. In a follow up study~\cite{GelisJV1}, we will address the
connections of our approach to kinetic
approaches~\cite{MuellS1,Jeon3}.

\section*{Acknowledgements}
We thank Robi Peschanski, Jochen Bartels and George Sterman for
enlightening discussions on the AGK rules, Sangyong Jeon for useful
discussions during the early stages of this work, and our colleagues
at Saclay and Brookhaven and Keijo Kajantie for many discussions on
related topics. RV's research was supported by DOE Contract
No. DE-AC02-98CH10886. FG wishes to thank the hospitality of TIFR,
Mumbai, and the support of CEPIPRA project No 3104-3.

\appendix

\section{Schwinger-Keldysh formalism}
\label{app:SK}
\begin{figure}[htbp]
\begin{center}
\resizebox*{!}{1cm}{\includegraphics{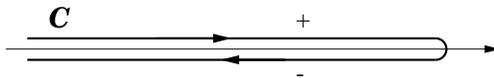}}
\end{center}
\caption{\label{fig:ctp}The closed time path used in the
Schwinger-Keldysh formalism.}
\end{figure}

Our intuitive discussion of cutting rules for computing probabilities
in field theories coupled to external sources, begun in
section~\ref{sec:vacuum} and continuing in
section~\ref{sec:distribution}, culminates in the formal expression of
these, in section~\ref{sec:Pn-1}, in terms of the Schwinger-Keldysh
vacuum-vacuum amplitudes. The Schwinger-Keldysh formalism was
developed in the 1960's in the context of many-body quantum field
theory \cite{Schwi1,Keldy1}. It is used primarily in thermal field
theory problems in order to calculate thermal averages of
operators~\cite{Bella1}. Since it may be less familiar to those
working outside this area, we will provide a brief review of the
formalism before proceeding to prove a result stated in
section~\ref{sec:Pn-1}.

The primary feature of the Schwinger-Keldysh formalism, relative to the
standard formulation of quantum field theory, is that the time flow of paths is
defined on a contour comprised of two branches (see the figure
\ref{fig:ctp}) wrapping the real axis. The upper branch,
denoted $+$, runs along the real axis in the positive direction while
the lower branch, denoted $-$, runs in the opposite direction. 
The aim is to compute the correlation functions
\begin{equation}
G^{\{\epsilon_1\cdots\epsilon_n\}}(x_1\cdots x_n)
\equiv
\big<0_{\rm in}\big|
P_{\cal C}\,\phi^{(\epsilon_1)}(x_1)\cdots\phi^{(\epsilon_n)}(x_n)
\big|0_{\rm in}\big>\; ,
\label{eq:SK-corr}
\end{equation}
where $P_{\cal C}$ denotes a path ordering of operators along the time
contour.  Operators are ordered from left to right by decreasing
values of their curvilinear abscissa on the contour, with operators
living on the lower branch placed to the left of operators living on
the upper branch. When restricted to the upper branch, it is
equivalent to the usual time ordering. Similarly, it is equivalent to
an anti-time ordering on the lower branch. In eq.~(\ref{eq:SK-corr}),
the indices $\epsilon_i$ can have the values $+/-$ indicating the
branch of the contour where the corresponding field is located. This
definition of the path ordering suggests that the correlator defined
in eq.~(\ref{eq:SK-corr}) can also be written as
\begin{equation}
G^{\{\epsilon_1\cdots\epsilon_n\}}(x_1\cdots x_n)
=
\big<0_{\rm in}\big|
\overline{T}\Big[\prod_{i|\epsilon_i=-}\phi^{(\epsilon_i)}(x_i)\Big]
          T \Big[\prod_{j|\epsilon_j=+}\phi^{(\epsilon_j)}(x_j)\Big]
\big|0_{\rm in}\big>\; .
\end{equation}
Note that the vacuum state used in order to define this correlation
function is the ``in'' vacuum state on both sides\footnote{This
  feature is a consequence of the fact that the thermal average of an
  operator is a {\bf trace} of the operator multiplied by the density
  matrix.}.
   
The perturbative calculation of the correlator in
eq.~(\ref{eq:SK-corr}) requires special Feynman rules. These are
obtained in the Schwinger-Keldysh formalism and their derivation is
analogous to that of regular time-ordered correlation functions. One
first expresses the Heisenberg fields in terms of the free field
$\phi_{\rm in}$ of the interaction picture. One can then rewrite
eq.~(\ref{eq:SK-corr}) as
\begin{equation}
G^{\{\epsilon_1\cdots\epsilon_n\}}(x_1\cdots x_n)
=
\big<0_{\rm in}\big|
P_{\cal C}\,\phi^{(\epsilon_1)}_{\rm in}(x_1)\cdots\phi^{(\epsilon_n)}_{\rm in}(x_n)\;
e^{\int_{\cal C}d^4x\,{\cal L}_{\rm int}(\phi_{\rm in}(x))}
\big|0_{\rm in}\big>\; .
\end{equation}
Note that the time integration in the exponential now runs over the
full contour ${\cal C}$. The perturbative expansion of this correlator
is obtained by expanding the exponential. This procedure differs from the 
usual Feynman rules in the following two ways:
\begin{itemize}
\item at each vertex, the time integration runs over the full contour ${\cal C}$
\item the free propagator connecting vertices is the
path-ordered product of two free fields
\begin{equation}
G^0(x,y)\equiv \big<0_{\rm in}\big|
P_{\cal C}\,\phi_{\rm in}(x)\phi_{\rm in}(y)
\big|0_{\rm in}\big>\; .
\label{eq:SK-prop}
\end{equation}
\end{itemize}
These rules are implemented by first breaking the time integration at
each vertex in two pieces. One corresponds to an integration over
times in the upper branch while the other integrates over the lower
branch. The vertices come in two varieties, $+$ vertices and $-$
vertices, and the indices must be summed over at each vertex. For
example, the contribution of the $-$ vertices is written as an
integration over the real axis in the positive direction modulo the
addition of a minus sign to the vertex of type $-$. This follows from
the fact that the integration runs in the negative direction on the
lower branch of ${\cal C}$.  Likewise, the propagator in
eq.~(\ref{eq:SK-prop}) is broken into four different propagators,
corresponding to where the two fields lie on the contour ${\cal C}$,
\begin{eqnarray}
&&
G^0_{++}(x,y)\equiv \big<0_{\rm in}\big|
T\,\phi_{\rm in}(x)\phi_{\rm in}(y)
\big|0_{\rm in}\big>\; ,
\nonumber\\
&&
G^0_{--}(x,y)\equiv \big<0_{\rm in}\big|
\overline{T}\,\phi_{\rm in}(x)\phi_{\rm in}(y)
\big|0_{\rm in}\big>\; ,
\nonumber\\
&&
G^0_{-+}(x,y)\equiv \big<0_{\rm in}\big|
\phi_{\rm in}(x)\phi_{\rm in}(y)
\big|0_{\rm in}\big>\; ,
\nonumber\\
&&
G^0_{+-}(x,y)\equiv \big<0_{\rm in}\big|
\phi_{\rm in}(y)\phi_{\rm in}(x)
\big|0_{\rm in}\big>\; .
\end{eqnarray}
Note that the Fourier transforms of these free propagators are
precisely those encountered previous in the derivation of cutting
rules (see eq.~(\ref{eq:Fourier-props})). The propagator
$G^0_{\epsilon\epsilon^\prime}$ connects a vertex of type $\epsilon$
to a vertex of type $\epsilon^\prime$.

In the discussion in section~\ref{sec:cutting}, the $\pm$ vertices,
propagators and sources were introduced as a formal trick to compute
the sum of the Feynman vacuum-vacuum amplitude and its complex
conjugate amplitude. We now understand these as arising naturally from
the Schwinger-Keldysh formalism which captures the physics of theories
in external fields/finite temperatures.  For example, an identity
(eq.~(\ref{eq:pm-constraint})) that we showed in
section~\ref{sec:cutting} is equivalent to a property of the
Schwinger-Keldysh formalism used in section \ref{sec:Pn-1}: {\sl the
  sum of all vacuum-vacuum diagrams is equal to one.} It remains true
even if the field is coupled to an external source, provided that
source is the same on the upper and lower branches of the contour
${\cal C}$.

The proof of this result in the Schwinger-Keldysh formalism is very
simple. Consider a connected vacuum-vacuum diagram, and integrate out
all but one of its vertices. The result is a function of a single
point we call $x$. This function, by virtue of not possessing external
legs, and being coupled to a source that has the same value on the $+$
and $-$ branches of the contour, has the same magnitude on both
branches of the contour. Therefore, when we sum over the $+/-$ type of
this last vertex, we get an exact cancellation because the $-$ index
has the opposite sign and the same magnitude as $+$ index. Thus all
{\bf connected} vacuum-vacuum diagrams are zero when the sources are
identical on both legs of the contour. Because the sum of {\bf all}
vacuum-vacuum diagrams, is the exponential of the sum of the connected
ones, it is therefore equal to unity.

\section{Solution of Lippman-Schwinger equation}
\label{app:LS}
The Lippman-Schwinger equation (\ref{eq:LS}) can be solved by
performing the following rotations of the propagators and field
insertions:
  \begin{align}
  &G_{\epsilon\epsilon^\prime}&\to&&&
  {\bs G}_{\alpha\beta}\equiv
\sum_{\epsilon,\epsilon^\prime=\pm}
U_{\alpha \epsilon}U_{\beta \epsilon^\prime}G_{\epsilon\epsilon^\prime}&&
  \nonumber\\
  &{\bs \sigma}^3_{\epsilon\epsilon^\prime}&\to&&&
  {\bs \Sigma}^3_{\alpha\beta}\equiv
\sum_{\epsilon,\epsilon^\prime=\pm}
U_{\alpha \epsilon}U_{\beta \epsilon^\prime}
{\bs \sigma}^3_{\epsilon\epsilon^\prime}&&
\label{eq:LS-trans}
\end{align}
with
\begin{equation}
U=\frac{1}{\sqrt{2}}\begin{pmatrix}1 & -1 \cr 1 & 1 \cr\end{pmatrix}\; .
\end{equation}
Under these rotations, the free matrix propagator and field insertion matrix become
\begin{equation}
  {\bs G}_{\alpha\beta}^0=
  \begin{pmatrix}
    0 & G_{_A}^0 \cr
    \cr
    G_{_R}^0 & G_{_S}^0 \cr
  \end{pmatrix}
  \quad,\qquad
  {\bs \Sigma}^3_{\alpha\beta}=
  \begin{pmatrix}
    0 & 1 \cr 1 & 0 \cr
  \end{pmatrix}\; .
\end{equation}

$G_{_R}^0$ and $G_{_A}^0$ are the free retarded and advanced propagators, and
$G_{_S}^0$ stands for the combination
\begin{equation}
G_{_S}^0\equiv G_{++}^0+G_{--}^0\; .
\end{equation}
Its Fourier transform is $G^0_{_S}(p)=2\pi\delta(p^2-m^2)$. In terms of these new propagators and vertices, the Lippmann-Schwinger equation has the same form as that of eq.~(\ref{eq:LS}), modulo the replacements of
eq.~(\ref{eq:LS-trans}). 
The main simplification follows from observing 
that the product ${\bs G}{\bs \Sigma}^3$ is the sum of a diagonal
matrix and a nilpotent matrix. This makes it easy to calculate  its
$n$-th power. In particular, we have\footnote{This formula is true
even if each propagator has different arguments. Therefore, it is also
true for the convolution of $n+1$ propagators, between which one
inserts the classical field.}
\begin{equation}
{\bs G}^0\left[{\bs\Sigma}^3{\bs G^0}\right]^n
=
\begin{pmatrix}
0 & \left[G_{_A}^0\right]^{n+1} \cr
\cr
\left[G_{_R}^0\right]^{n+1} & 
\sum_{i=0}^n\left[G_{_R}^0\right]^{i} G_{_S}^0 \left[G_{_A}^0\right]^{n-i}
\cr
\end{pmatrix}
\; .
\end{equation}
Note that the off-diagonal equations mean that in the new basis
the Lipp\-mann-Schwin\-ger equations do not mix the retarded and advanced propagators.

The result above can be easily inverted to obtain the propagators of the Schwinger--Keldysh formalism. 
One obtains\footnote{One can check that this solution satisfies the following properties
\begin{eqnarray}
&&
G_{++}+G_{--}=G_{-+}+G_{+-}
\nonumber\\
&&
G_{++}-G_{+-}=G_{_R}\; ,\quad G_{++}-G_{-+}=G_{_A}\; .
\nonumber
\end{eqnarray}
 }~
\begin{eqnarray}
&&
G_{-+}=G_{_R} G_{_R}^0{}^{-1} G_{-+}^0 G_{_A}^0{}^{-1}G_{_A}
\nonumber\\
&&
G_{+-}=G_{_R} G_{_R}^0{}^{-1} G_{+-}^0 G_{_A}^0{}^{-1}G_{_A}
\nonumber\\
&&
G_{++}=\frac{1}{2}\left[
G_{_R} G_{_R}^0{}^{-1} G_{_S}^0 G_{_A}^0{}^{-1}G_{_A}
+(G_{_R}+G_{_A})\right]
\nonumber\\
&&
G_{--}=\frac{1}{2}\left[
G_{_R} G_{_R}^0{}^{-1} G_{_S}^0 G_{_A}^0{}^{-1}G_{_A}
-(G_{_R}+G_{_A})\right]\; .
\label{eq:G1}
\end{eqnarray}
As one can see, all these propagators resumming the effect of the
classical background field $\phi_c$ can be expressed in terms of the
retarded and advanced propagator in the background field.  This is a
useful property because retarded propagators can be obtained by
solving a simple partial differential equation with retarded boundary
conditions. (Advanced propagators can be obtained from the retarded ones
by permuting the endpoints.)

\section{Reciprocal of eqs.~(\ref{eq:agk-ident})}
\label{app:inv-agk}
It is also instructive to see what the AGK identities, as given in
eqs.~(\ref{eq:agk-ident}), imply for the distribution of
multiplicities of cut subdiagrams in eq.~(\ref{eq:regg-P}). Let us
assume the following generic form for the probability ${\cal R}_p$ of
having $p$ cut subdiagrams~:
\begin{equation}
{\cal R}_p\equiv R(\eta)\;\frac{1}{p!}\,r_p\,\eta^p\; ,
\end{equation}
where $\eta$ is a parameter that counts the number of subdiagrams (in
the situation studied in section \ref{sec:reggeon}, the quantity
$a/g^2$ played this role). The prefactor $R(\eta)$ is determined from
the requirement that the sum of these probabilities be unity,
\begin{equation}
R(\eta)=\left[\sum_{p=0}^{+\infty}\frac{1}{p!}\,r_p\,\eta^p\right]^{-1}\; ,
\end{equation}
and the coefficients $r_p$ must be positive in order to have positive
probabilities. Moreover, we can assume without loss of generality that
$r_0=1$. The ${\cal R}_{p,q}$ that appear in eqs.~(\ref{eq:agk-ident})
are obtained by expanding the prefactor $R(\eta)$ to order $q-p$, and
are thus completely determined from the coefficients $r_p$. 

The relations (\ref{eq:agk-ident}) are very complicated non-linear
relations when expressed in terms of the coefficients $r_p$, because
of the prefactor $R(\eta)$. In order to investigate their
consequences, it is much easier to introduce the function
\begin{equation}
G(x)\equiv \sum_{p=0}^{+\infty}\frac{1}{p!}\,r_p\,(\eta x)^p\; .
\end{equation}
Note that the prefactor $R(\eta)$ is nothing but
$G^{-1}(1)$. Therefore, if we consider the following generating
function for the probabilities ${\cal R}_p$,
\begin{equation}
H(x)\equiv \sum_{p=0}^{+\infty}{\cal R}_p\,x^p\; ,
\label{eq:def-H}
\end{equation}
it is related to $G(x)$ via~:
\begin{equation}
H(x)=G(x)/G(1)\; .
\end{equation}
Recalling now that
\begin{equation}
{\cal R}_p=\sum_{q=p}^{+\infty}{\cal R}_{p,q}\; ,
\end{equation}
and using eqs.~(\ref{eq:agk-ident}), we have the following relations
for the successive derivatives of $H(x)$ at $x=1$~:
\begin{equation}
H^{(1)}(1)={\cal R}_{1,1}\;,\quad
H^{(2)}(1)=2!\,{\cal R}_{2,2}\;,\quad\cdots\quad
H^{(p)}(1)=p!\,{\cal R}_{p,p}\; .
\end{equation}
Therefore, we can rewrite the generating function $H(x)$ as 
\begin{equation}
H(x)=\sum_{p=0}^{+\infty}{\cal R}_{p,p}(x-1)^p = \sum_{p=0}^{+\infty}
\frac{1}{p!}\,r_p\,(x-1)^p=G(x-1)\; .
\end{equation}
In order to obtain the second equality, we have used the fact that
$R_{p,p}$ is obtained by keeping only the order 0 in the expansion of
the prefactor $R(\eta)$, which is equal to unity from our choice to
have $r_0=1$. Therefore, the AGK identities, eqs.~(\ref{eq:agk-ident}),
imply that
\begin{equation}
\forall x\; ,\quad G(x)=G(x-1)G(1)\; .
\label{eq:agk-func}
\end{equation}

One sees immediately that the exponentials, $G(x)=\exp(\alpha x)$, are
trivial solutions of this functional relation. Note that this solution
corresponds to a Poisson distribution for the probabilities ${\cal
R}_p$. We are now going to argue that this obvious solution is in fact
the only solution which is compatible with the positivity of the
coefficients $r_p$. Let us factor out of $G(x)$ the exponential
behavior it may contain by writing
\begin{equation}
G(x)=g(x)\,e^{\alpha x}\; ,
\end{equation}
where $g(x)$ is a function whose growth at infinity is bounded by a
polynomial. The functional equation~(\ref{eq:agk-func}), when
rewritten in terms of $g(x)$, implies
\begin{equation}
\forall x\; , \quad g(x+1)=g(x)g(1)\; ,
\label{eq:agk-func-1}
\end{equation}
and in particular
\begin{equation}
\forall n\in{\mathbbm Z}\; ,\quad g(n)=g(0) g^n(1)\; .
\end{equation}
Since $g(x)$ cannot grow faster than a polynomial at infinity, we must
therefore have $g(1)=1$, and eq.~(\ref{eq:agk-func-1}) becomes
\begin{equation}
\forall x\; ,\quad g(x+1)=g(x)\; .
\end{equation}
Hence the function $g(x)$ must be a periodic function of period
unity. This is where the positivity of the $r_p$'s plays a role: it
implies that all the successive derivatives of $G(x)$ are positive for
$x>0$. In terms of $g(x)$, the $n$-th derivative of $G(x)$ reads
\begin{equation}
G^{(n)}(x)=e^{\alpha x} \;\sum_{p=0}^nC_n^p\;\alpha^{n-p}\,g^{(p)}(x)\; ,
\label{eq:agk-ineq}
\end{equation}
where the $C_n^p\equiv n!/p!(n-p)!$ are the binomial coefficients. We
must therefore find a periodic function $g(x)$ such that
\begin{equation}
\forall n\in{\mathbbm N}\; ,\quad \forall x\; ,\quad 
\sum_{p=0}^nC_n^p\;\alpha^{n-p}\,g^{(p)}(x) >0\; ..
\end{equation}
Let us write the periodic function $g(x)$ as a Fourier series,
\begin{equation}
g(x)\equiv \sum_{l\in {\mathbbm Z}} g_l \; e^{2i\pi l x}\; ,
\end{equation}
where $g_{-l}=g_l^*$ since $g(x)$ is real when $x$ is real. The left
hand side in the inequality (\ref{eq:agk-ineq}) can be written as
\begin{equation}
g_0\,\alpha^n+2\sum_{l=1}^{+\infty}
\left|g_l\right|\; (\alpha^2+4\pi^2 l^2)^{n/2} \;\cos(2\pi l x+n\phi_l)\; ,
\label{eq:agk-ineq-2}
\end{equation}
where we parameterize $g_l(\alpha+2i\pi l)\equiv
\left|g_l\right|\sqrt{\alpha^2+4\pi^2 l^2} \exp(i\phi_l)$\; . We see
that for any $n$ and $l$, we can choose a value of $x$ such that the
cosine equals $-1$. Moreover, by going to sufficiently large $n$, we
can make $(\alpha^2+4\pi^2 l^2)^{n/2}$ arbitrarily large compared to
$\alpha^n$. Therefore, the values of $\left|g_l\right|$ such that
eq.~(\ref{eq:agk-ineq-2}) is always positive have to be chosen
arbitrarily small. Therefore, we have proved that the only choice of
the Fourier coefficients which is compatible with the positivity of
the probabilities ${\cal R}_p$ is
\begin{equation}
\forall l\in{\mathbbm Z}^*\; ,\quad g_l=0\; .
\end{equation}
Thus, the function $g(x)$ is constant and equal to $1$ (from
$g(1)=1$), and the only functions $G(x)$ that obey the functional
relation (\ref{eq:agk-func}) are the exponentials $G(x)=\exp(\alpha
x)$. We have therefore proved that the only distributions ${\cal R}_p$
that lead to the AGK cancellations are Poisson distributions.

\section{Discussion of the cancellations in eq.~(\ref{eq:agk-rhs2})}
\label{app:cancellations}
We will here study the cancellations in eq.~(\ref{eq:agk-rhs2}) more
explicitly. Note first that the terms in $g^{-2}\sum_r rb_r$ must
correspond to the terms in the r.h.s. of eq.~(\ref{eq:agk-rhs2})
topology by topology. For the simplest topology that appears in
$b_1/g^2$, only 1-particle cuts are allowed, and it is trivial to
check the identity of eq.~(\ref{eq:agk-rhs2}). The first non-trivial
topology is the ``star'' diagram that appears in $b_1/g^2$ and in
$b_2/g^2$. Its contribution to the left hand side of
eq.~(\ref{eq:agk-rhs2}) is \setbox1=\hbox to
8cm{\resizebox*{8cm}{!}{\includegraphics{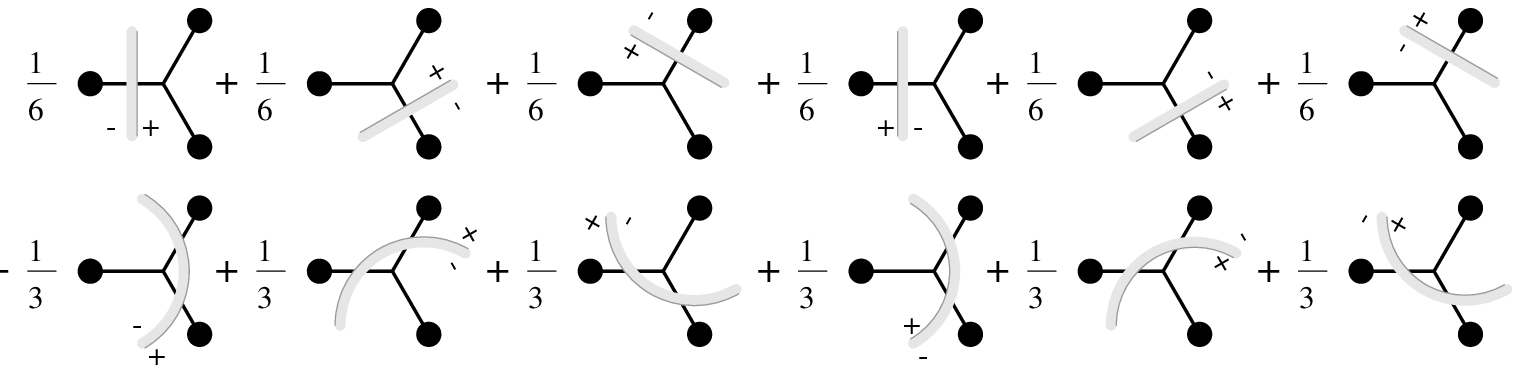}}}
\begin{equation}
\frac{1}{g^2}\sum_{r}\left.rb_r\right|_{_{LO}}
=\quad\;\raise -14.5mm\box1
\label{eq:rbr}
\end{equation}
On the other hand, the contribution of this topology to the right hand
side of eq.~(\ref{eq:agk-rhs2}) is comprised of the terms
\setbox1=\hbox to
8cm{\resizebox*{8cm}{!}{\includegraphics{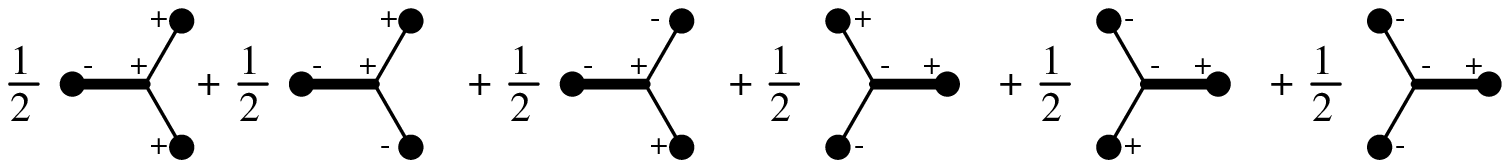}}}
\begin{equation}
\quad\;\raise -5.5mm\box1
\label{eq:agk-rhs3}
\end{equation}
Note that we have discarded two vanishing terms, in which a vertex of
type $+$ (resp. $-$) was completely surrounded by sources of type $-$
(resp. $+$), because such a configuration is forbidden by energy
configuration (because of the $\theta(\pm p_0)$ factors in the Fourier
transform of the propagators $G_{-+}$ and $G_{+-}$). At this point, it
is a trivial matter of inspection to check that eqs.~(\ref{eq:rbr})
and (\ref{eq:agk-rhs3}) are identical. One should stress that it was
crucial to weight the various $r$-particle cuts $b_r/g^2$ by the
multiplicity $r$ on the cut.

One can also see that the equivalence stated in
eq.~(\ref{eq:agk-rhs2}) between the method of calculating
$\big<n\big>$ from the classical field $\phi_c$ and the direct method
implies that $b_2\not=0$ (in other words, this equivalence fails if
one disregards the 2-particle cuts in
eq.~(\ref{eq:rbr})). Incidentally, this proves that the distribution
of produced particles is not a Poisson distribution, according to the
discussion at the end of section \ref{sec:distribution}.

\bibliographystyle{unsrt}

\end{document}